\documentclass[twocolumn,%
nofootinbib,
 amsmath,amssymb,
 aps,
prd,
]{revtex4-2}
\newcommand{\be}{\begin{equation}}
\newcommand{\ee}{\end{equation}}
\newcommand{\bea}{\begin{eqnarray}}
\newcommand{\eea}{\end{eqnarray}}
\newcommand{\bwt}{\begin{widetext}}
\newcommand{\ewt}{\end{widetext}}

\newcommand{\bi}{\begin{itemize}}
\newcommand{\ei}{\end{itemize}}

\usepackage{xcolor}
\usepackage{braket}
\usepackage{setspace}

\usepackage{graphicx}
\usepackage{dcolumn}
\usepackage{bm}

\begin{document}


\title{The primordial non-Gaussianities\\ for non-singular Horndeski cosmologies}

\author{Y.~Ageeva}
 \email{ageeva@inr.ac.ru}
\affiliation{Institute for Nuclear Research of
         the Russian Academy of Sciences,\\  60th October Anniversary
  Prospect, 7a, 117312 Moscow, Russia,\\
Department of Particle Physics and Cosmology, Physics Faculty, M.V. Lomonosov Moscow State University, \\Leninskie Gory 1-2, 119991 Moscow, Russia}
\author{M. Kotenko}
 \email{max.kotenko@yandex.ru}
 \affiliation{Institute for Nuclear Research of
         the Russian Academy of Sciences,\\  60th October Anniversary
  Prospect, 7a, 117312 Moscow, Russia,\\Department of Particle Physics and Cosmology, Physics Faculty, M.V. Lomonosov Moscow State University, \\Leninskie Gory 1-2, 119991 Moscow, Russia}
\author{P.~Petrov}
 \email{petrov@inr.ac.ru,p669590371@ibs.re.kr}
\affiliation{Cosmology, Gravity, and Astroparticle Physics Group,\\
Center for Theoretical Physics of the Universe,\\
Institute for Basic Science (IBS), Daejeon, 34126, Korea}

\date{\today}

\begin{abstract}
We consider a novel Bounce Universe model constructed within the framework of Horndeski gravity. We have analyzed the bispectrum of primordial scalar perturbations and evaluated the corresponding non-Gaussianities for this specific model. 
The non-linear parameter $f_{\text{NL}}$
was computed for various wave vector configurations, including: (i) the local configuration, (ii) the equilateral configuration, and (iii) the enfolded configuration, which is a linear combination of the equilateral and orthogonal shapes. 
We demonstrate that the observational constraints on the scalar spectrum index tilt and the scalar-to-tensor ratio, together with the unitarity bounds, ensure that the bounds from non-Gaussianities are trivially satisfied. Therefore, this particular Bounce Universe scenario is fully viable in light of modern observational data.

\end{abstract}

\maketitle


\section{Introduction}
\label{sec:intro}

The inflation theory \cite{Starobinsky:1980te,Guth:1980zm,Sato:1981qmu,Linde:1981mu,Albrecht:1982wi} is a cornerstone of the modern cosmology. It  solves  many problems that the Big Bang theory alone could not overcome. Nonetheless, there is no single inflation model, instead there is a plethora of different implementations of this idea and even more alternatives to it, which all more or less satisfy the current observable data, for instance \cite{Planck:2018vyg,BICEP:2021xfz}.
Thus, the constraints from the power spectrum alone are not enough to rule out the ones that do not describe our Universe. To this end, the higher order correlation functions are needed. In the coming years with the increase of accuracy of the measurements of non-Gaussianity which is described by the three-point correlation functions, it would be possible to further constrain or even to completely discard many inflationary scenarios and alternatives to it. 

Another problem of the inflation theory together with the seeming arbitrariness of the choice of the models is that there necessarily exists an initial singularity which leads to geodesic incompleteness \cite{Penrose:1964wq,Hawking:1966vg,Borde:1993xh,Borde:1996pt,Borde:2001nh,Yoshida:2018ndv}. To fix the problem of singularity in the framework of the classical field theory it is necessary \footnote{Actually, one can choose other ways to build non-singular models, for example, in Refs.~\cite{Arefeva:2011qvf,Arefeva:2011yua,Silva:2015qna,Silva:2020bnn,Silva:2023ieb} and others, authors use the holography and/or different string fields theory features to solve different gravitational singularities.} to violate the appropriate energy condition \cite{Hawking:1973uf,Tipler:1978zz}, and in the context of general relativity (GR) one usually addresses \footnote{If at some period of evolution gravity is not described by the standard GR, then a more general condition should be considered $R_{\mu\nu}n^{\mu}n^{\nu}\geq0$, see Ref.~\cite{Tipler:1978zz}.} to the Null Energy Condition (NEC):
\begin{equation*}
    T_{\mu\nu}n^\mu n^\nu>0,
\end{equation*}
where $n^{\mu}$ is a light-like vector, and $ T_{\mu\nu}$ is a stress-energy tensor.
This violation is necessary since according to the Penrose theorem \cite{Penrose:1964wq} the appearance of the singularity would be unavoidable otherwise. Note, that applied to the modern Universe the NEC makes it so that the Hubble parameter can not grow. If the Hubble parameter is growing currently, it would show that either the NEC is violated for the dark energy or that the GR is untrue for the modern cosmological scales \cite{Rubakov:2014jja}. Therefore, it is worth to consider the modified theories of gravity for the construction of non-singular cosmological scenarios of the early Universe. 

Unfortunately, the realization of such models within the framework of the classical field theory is burdened by instabilities. One of the most popular  theories that can violate NEC in a stable way without pathologies was proposed by G.~W.~Horndeski \cite{Horndeski:1974wa}. It is the most general formulation of a scalar-tensor theory of gravity. Its Lagrangian contains the derivatives of the second order, but the resulting field equations are still of the second order, which is necessary to avoid pathological degrees of freedom. With the development of this theory, many novel non-singular cosmological scenarios were proposed. For instance, there are the models of the Universe with a bounce that were discussed in the review paper \cite{Kobayashi:2019hrl} and in others \cite{Battefeld:2014uga,Lehners:2008vx,Brandenberger:2016vhg,Akama:2025ows}, as well as a Galilean Genesis \cite{Creminelli:2010ba,Creminelli:2012my,Nishi:2015pta, GilChoi:2025hbs}. 

The cosmological models with early non-singular epoch usually suffer from different kinds of pathologies, one of them is an Ostrogradsky instability \cite{Ostrogradsky:1850fid}. However, as it was mentioned above, the Horndeski theory by construct allows one to avoid them entirely. Still, such a modified gravity has other problems. The one that is important to us in this work is that the models built within a framework of Horndeski theories can not be trivially extended over the whole time interval $-\infty<t<+\infty$, since it leads to instabilities. This property is proved by the so-called ``no-go theorem'', which is true for Horndeski theories in a quite general sense  \cite{Kobayashi:2016xpl,Libanov:2016kfc}. 
This theorem states that under reasonable conditions of $a>\text{const }(>0)$  and of the smoothness of the Lagrangian functions that depend on $\phi$ and Hubble parameter $H$, the following integrals coming from the considerations about the perturbations around the flat Friedmann-Lemaître-Robertson-Walker (FLRW) background (i.e. $ds^2=-dt^2+a^2(t)\delta_{ij}dx^idx^j$) are diverging:
\begin{align*}
     \int_{-\infty}^ta(t)(\mathcal{F}_T+\mathcal{F}_S)dt=\infty, \\\int_t^{+\infty}a(t)(\mathcal{F}_T+\mathcal{F}_S)dt=\infty,
\end{align*}
where $\mathcal{F}_T$ and $\mathcal{F}_S$ are time-dependent coefficients of the quadratic action of the tensor and the scalar perturbations, respectively. 
The presence of such divergences is problematic for the model since it leads to instabilities. But there is a way to circumvent this theorem by using the theories beyond Horndeski, \cite{Gleyzes:2014dya} and the so-called Degenerate High Order Scalar-Tensor theories (DHOST) theories \cite{Langlois:2015cwa}. Unfortunately, they have their own problems such as superluminal speeds in presence of the matter field, see Refs.~\cite{Kolevatov:2017voe,Mironov:2020pqh,Mironov:2022ffa,Mironov:2024yqa,Mironov:2024pjt}.

In this paper the no-go theorem was circumvented by tending the coefficients of the quadratic action for scalar and tensor perturbations to zero in the distant past:
\begin{align*}
&\mathcal{S}_{hh} =\int dtd^{3}x\frac{a^{3}}{8}\left[\mathcal{G}_{T}\Big(\frac{\partial h_{ij}}{\partial t}\Big)^{2}-\frac{\mathcal{F}_{T}}{a^{2}}h_{ij,k}h_{ij,k}\right],  \\
&\mathcal{S}_{\zeta\zeta} =\int dtd^{3}xa^{3}\left[\mathcal{G}_{S}\Big(\frac{\partial\zeta}{\partial t}\Big)^{2}-\frac{\mathcal{F}_{S}}{a^{2}}\zeta_{,i}\zeta_{,i}\right], 
\end{align*}
i.e. $\mathcal{G}_T,\mathcal{F}_T,\mathcal{G}_S,\mathcal{F}_S\to0$ as $t\to-\infty$. Since the coefficients $\mathcal{G}_T,\mathcal{F}_T,\mathcal{G}_S,\mathcal{F}_S$ are effectively analogous to $M_{\text{Pl}}^2$, this leads to the strong gravity in the past. This causes a pathologically fast growth of quantum fluctuations, which makes it impossible to describe the model within the framework of the classical theory. But in the \textit{Bounce Universe} model,\footnote{The details of construction of this specific cosmological scenario can be found in Ref.~\cite{Ageeva:2022asq}, Appendix F. There, the authors give an explicit numerical example of stable and subluminal evolution from
contraction stage to bounce and then to kination epoch.} that is used in this paper this does not hold as a truth. Indeed, in Ref.~\cite{Ageeva:2022asq} such the conditions were applied on the parameters of the model, so that became possible to describe the evolution of the Universe within the classical frameworks at all times. It is possible to achieve since in that model of Ref.~\cite{Ageeva:2022asq} the classical energy scale stays lower than the  quantum strong coupling scale.

The healthy construction of early-time cosmology, i.e. avoiding all instabilities and other pathologies like strong coupling regime imposes significant restrictions on the model parameters. But even stricter constraints can be also obtained by considering the latest observational data \cite{Planck:2018vyg,BICEP:2021xfz}.  Along with the spectral index $n_S$ in power spectra and the $r$-ratio of the amplitudes of scalar and tensor perturbations, as it was stated before in Ref.~\cite{Ageeva:2022asq}, the non-Gaussianity of the primordial scalar perturbations of the metric may also be used to choose the fitting models of inflation or their alternatives. Therefore by calculating the non-Gaussianity non-linear parameter $f_{\text{NL}}$ for chosen model we would be able to further constrain its parameters or to even outright discard it. 

In this paper we calculate the non-Gaussianity non-linear parameters $f_{\text{NL}}$ for various forms or momenta configurations such as equilateral, where $k_1=k_2=k_3=k$; local, where $k_1\to0,k_2\simeq k_3$; and enfolded with $k_1=k_2+k_3$. The results are in accordance with the observational constraints, which were acquired by \textit{Planck collaboration}, namely: $f_{\mathrm{NL}}^{\mathrm{local}}=-0.9\pm5.1,  f_{\mathrm{NL}}^{\mathrm{equil}}=-26\pm47$, and  $f_{\mathrm{NL}}^{\mathrm{enfold}}=6\pm30.5$, (all with $68\;\%\; \mathrm{CL}$, statistical), see Ref.~\cite{Planck:2018vyg}.
The parameter $f_{\mathrm{NL}}^{\mathrm{enfold}}$ is given by $f_{\mathrm{NL}}^{\mathrm{enfold}}=(f_{\mathrm{NL}}^{\mathrm{equil}}-f_{\mathrm{NL}}^{\mathrm{ortho}})/2$, what is similar to  Ref.~\cite{Meerburg:2009ys}. 

However, in our case of contracting Universe, it is impossible to directly apply method from the Refs.~\cite{DeFelice:2011zh,DeFelice:2013ar} in order to evaluate non-Gaussianity. Indeed, in Refs.~\cite{DeFelice:2011zh,DeFelice:2013ar} one evaluates non-linear parameters $f_{\text{NL}}$ in the case of slow-roll inflation, however our model is contracting Universe with the subsequent Bounce. Therefore, we move to the Einstein frame, where in our setup we have inflationary model, see Ref.~\cite{Ageeva:2022asq}. For our model the effective Planck mass vanishes in asymptotic past, thus strictly speaking the transition between frames becomes singular, when time goes to minus infinity. Nevertheless, for any finite negative time one can perform conformal transformation and move to Einstein frame. In other words, in our model the Jordan and Einstein frames are not equivalent, however for any finite time it is fully legitimate to make the transition between frames.
As it was mentioned before, after the moving to the Einstein frame our model becomes the power-law G-inflation \cite{Ageeva:2022asq}. The Lagrangian in the Einstein frame is particularly easier than in Jordan frame, so moving to Einstein frame simplifies the evaluations sufficiently. Moreover, the G-inflation in Einstein frame is a slow-roll model. That is why, in the same manner as in Ref.~\cite{DeFelice:2013ar}, we use the expansion of all $f_{\text{NL}}$ with respect to slowly varying parameters up to the leading non-zero order. In the framework of this expansion, we have explicitly shown that the standard Maldacena's consistency relation \cite{Maldacena:2002vr} holds for $f_{\text{NL}}^{\text{local}}$. It is known, that for Bounce models consistency relation could break, see, for instance, Ref.~\cite{Chowdhury:2015cma}. However, our model has \textit{non-minimal coupling between gravity and scalar field}, therefore our current results are in agreement with Ref.~\cite{Nandi:2019xag}, where it was shown that consistency relation may hold for Bouncing Universe if one introduces  mentioned non-minimal coupling. Actually, consistency relation has a quite generic nature and holds for many other cosmological models. Indeed, it is satisfied for many inflationary models, see for example Refs.~\cite{Creminelli:2004yq,Seery:2005wm,Chen:2006nt,Cheung:2007sv,Ganc:2010ff,Renaux-Petel:2010paw,Kundu:2014gxa,Kundu:2015xta,Bravo:2017wyw}. Furthermore, we should note, that if the Bounce Universe scenario is described by a single-field theory, the adiabatic initial condition is imposed, and both quadratic and cubic actions are conformally equivalent to inflation, then the Maldacena’s consistency relation should hold as well. The latter is particularly the case for the Ref.~\cite{Akama:2025ows}, where author studies another variant of Bounce with scale-invariant scalar and tensor power spectra and analyses non-Gaussianities wherein. Turning back to our model, for the evaluated parameters $f_{\text{NL}}^{\text{equil}}$ and $f_{\text{NL}}^{\text{enfold}}$ as a functions of model parameters we have applied experimental constraints on non-Gaussianities from \textit{Planck} \cite{Planck:2018vyg}. We have found that this bounds indeed imposes nontrivial constraints on model parameters.  We explicitly show, how new bounds from non-Gaussianities relates with the rest of different conditions which we have studied earlier in Ref.~\cite{Ageeva:2022asq}: the experimental values of scalar spectral tilt, amplitude of scalar perturbations, $r$-ratio value,  avoiding of Belinsky--Khalatnikov--Lifshitz (BKL) problem \cite{Lifshitz:1963ps, Belinsky:1970ew,Belinskii:1972sg} and instabilities during the whole evolution. It turns out that the observational constraints for power spectrum tilt $n_S$ and $r$-ratio together with the theoretical unitarity bounds ensure that constraints from non-Gaussianities hold. The latter particularly means that the analysis of non-Gaussianities does not close the considered cosmological model of contracting Universe with the subsequent bounce from \cite{Ageeva:2022asq}, as well as does not lead to any new bounds on the parameter space.

The paper is organized as follows. In Section \ref{sec:model} we give a description and properties of the scenario of the  contracting Universe within the general action of the Horndeski theory: in Sec.~\ref{sec:prop}, in terms of the Arnowitt-Deser-Misner (ADM) formalism we show the Ansatz for the Lagrangian and conditions on its parameters which together admit non-pathological (at all times) contraction without strong coupling regime in the past as well as recall that it is also possible to generate the primordial perturbations with the features being in agreement with observational data \cite{Planck:2018vyg,BICEP:2021xfz}. In Sec.~\ref{sec:param} we provide the covariant form of the Lagrangian as well; the latter is valuable when choose the physical parameters of the model. In the same Section we move to the Einstein frame and show, that there we have the slow-roll G-inflation, deriving corresponding slowly varying parameters. 
In Sec.~\ref{sec:nongauss} we briefly review the concept of non-Gaussianity itself and the method of their evaluation. 
The Sec.~\ref{sec:eval non gauss} is dedicated to the calculation of the non-Gaussianity parameters. Finally, we provide the dependence of $f_{\mathrm{NL}}$ on our model parameters $\mu$ and $\chi$ for different combinations of momenta (wave numbers), what is the main result of our work. In the same Section the proof of Maldacena's consistency relation for local form of non-Gaussianity can be found; also the comparison of our results for non-linear parameters with the latest observational data gives the new constraints on $\mu$ and $\chi$. We conclude in Section~\ref{sec:summary}. The paper has three Appendices as well. The Appendix~\ref{app:sec_ord} collects the expressions for the coefficients from second order action for scalar and tensor perturbations. The explicit solution of linearized equation of motion for scalar perturbation is shown in Appendix~\ref{app:pert}. 
Finally,  in Appendix~\ref{app:slowroll} we explicitly show, that slow-roll approximation is inapplicable for contracting Universe model, therefore we move to the Einstein frame with slow-roll inflation and use the formulas for non-Gaussianities from Ref.~\cite{DeFelice:2013ar}.

\section{Generalities}
\label{sec:model}
In this paper we consider the same model of power-law contraction in the framework of Horndeski theory as in Ref.~\cite{Ageeva:2022asq}. The corresponding action reads
    \begin{align}
      \cal S=&\int d^4x \sqrt{-g}
      \big\{ G_2(\phi, X)-G_3(\phi, X)\Box \phi\nonumber\\
      &+ G_4(\phi,X)R + G_{4X}\big[(\Box \phi)^2 - (\nabla_{\mu}\nabla_{\nu}\phi)^2\big]\big\}\;,
    \label{Hor_L}\\
        X =& -\frac{1}{2}g^{\mu\nu}\partial_{\mu}\phi\partial_{\nu}\phi\;,\nonumber
    \end{align}
where 
$\Box \phi = g^{\mu\nu} \nabla_\mu \nabla_\nu \phi$ 
and $(\nabla_{\mu}\nabla_{\nu}\phi)^2 = \nabla_{\mu}\nabla_{\nu}\phi \, \nabla^{\mu}\nabla^{\nu}\phi$, and 
$R$ is the Ricci scalar; also we use $G_{X} \equiv \partial G/ \partial X$ and $G_{\phi} \equiv \partial G/ \partial \phi$. We stick to the following metric
signature: $(-,+,+,+)$. While the general Horndeski theory involves four arbitrary functions -- $G_{2,3,4,5}$,
the chosen Lagrangian \eqref{Hor_L} has only three ones.

For our purposes it is much more  convenient to work in the ADM formalism \cite{Kobayashi:2019hrl}.
There, the metric is written as
 \begin{equation*}
     ds^2=-N^2 d\hat{t}^2 +  
        \gamma_{ij}\left( dx^i+N^i d\hat{t}\right)\left(dx^j+N^j d\hat{t}\right) \;,
    \end{equation*}
with  $\gamma_{ij}$ being the three-dimensional metric,
$N$ is the \textit{lapse} function and  $N_i=\gamma_{ij}N^j$ 
is the \textit{shift}  vector.
We hold the same notations as in Ref.~\cite{Ageeva:2022asq} and denote the general time variable by $\hat{t}$, while reserving the
notation $t$ for cosmic time in the further calculations.
By choosing field $\phi$ depending on $\hat{t}$
only, i.e. $\phi=\phi(\hat{t})$ (the unitary gauge), one can rewrite the
action as follows
\begin{align}
\label{adm_lagr}
        &\mathcal{S} = \int d^4x \sqrt{-g}    \Big[ A_2 (\hat{t}, N) + A_3 (\hat{t}, N) K \nonumber\\
        &+  A_4 (\hat{t}, N)
        (K^2 - K_{ij}^2) + B_4 (\hat{t}, N) R^{(3)}  \Big]\; \text{,}
\end{align}
where
\begin{equation*}
    A_4(\hat{t},N) = - B_4(\hat{t},N) - N\frac{\partial B_4(\hat{t},N)}{\partial N}\;,
\end{equation*}
and $^{(3)} R_{ij}$ is the Ricci tensor built by $\gamma_{ij}$, $\sqrt{-g} = N\sqrt{\gamma}$, $K= \gamma^{ij}K_{ij}$, $^{(3)} R = \gamma^{ij} \phantom{0}^{(3)} R_{ij}$ and
\begin{align*}
      K_{ij} &\equiv\frac{1}{2N}\left(\frac{d\gamma_{ij}}{d\hat{t}}
      -\,^{(3)}\nabla_{i}N_{j}-\;^{(3)}\nabla_{j}N_{i}\right) ,
\end{align*}
is extrinsic curvature of hypersurfaces $\hat{t}=\mbox{const}$. The Lagrangian functions in the covariant and ADM formalisms are related as
~\cite{Gleyzes:2014dya, Gleyzes:2013ooa, Fasiello:2014aqa}
\begin{align}
    G_2 =  A_2 - 2XF_{\phi} \text{,} \;  \;\;\;\;
    G_3 = - 2XF_X - F \text{,}  \;  \;\;\;
    G_4 = B_4 \;\text{,}  
\label{jan23-22-1}
\end{align}
where one also has for $N$ and $X$ 
\begin{align}
      N^{-1} d\phi/d\hat{t} = \sqrt{2X}\;,
    \label{jan25-22-30}
\end{align}
    and
\begin{equation}
\label{F}
    F_X = - \frac{A_3}{\left(2X\right)^{3/2}} - \frac{B_{4\phi}}{X}\; \text{.}
\end{equation}

To describe FLRW background solutions and perturbations \footnote{Note, that in this paper we consider the  perturbations at linear level only, however one can go further and study the 2nd-order perturbation beyond the linear one to include nonlinear effects, see e.g.~Ref.\cite{Wang:2017krj,Wang:2019zhj,Wang:2023nsj}.  }, one writes 
\begin{subequations}
\label{jul17-22-2}
\begin{align}
        N &=N_0(\hat{t}) (1+\alpha)\;,\quad
        N_{i} =\partial_{i}\beta +  N^T_i\;,
        \\
        \gamma_{ij} &=a^{2}(\hat{t}) \Big(\text{e}^{2\zeta}(\text{e}^{h})_{ij} + \partial_i\partial_j Y 
        + \partial_i W^T_j + \partial_j W^T_i\Big) \; ,
\end{align}
\end{subequations}
        where
        $a(\hat{t})$ and $N_0(\hat{t})$ are background solutions, which can be found from equations of motion for spatially flat FLRW background \cite{Kobayashi:2011nu}
\begin{subequations}
\label{eoms}
    \begin{eqnarray}
         &  (NA_2)_N + 3NA_{3N}H + 6N^2(N^{-1}A_4)_N H^2 = 0\;,\\
      &  A_2
        -6A_4H^2-\frac{1}{N}\frac{d}{d\hat{t}}\left( A_3+4A_4H \right) = 0 \; ,
    \end{eqnarray}
\end{subequations}
where $A_{N} \equiv \partial A/ \partial N$, and $H = (Na)^{-1} (da/d\hat{t}$) is the Hubble parameter and where not to encumber the formulas, we denote the background value of lapse as $N$ instead of $N_0$.  
Next, we also have $\partial_i N^{Ti}=0$ and
         \begin{align*}
        &(\text{e}^h)_{ij} =\delta_{ij}+h_{ij}+\frac{1}{2}h_{ik}h_{kj}+\frac{1}{6}h_{ik}h_{kl}h_{lj}+
        \cdots\;, \\ &h_{ii}=0\;, \quad
        \partial_{i}h_{ij}=0 \; .
    \end{align*}  
The residual gauge freedom is fixed
by setting $Y = 0$ and $W^T_i = 0$.
It is also known, that the variables
$\alpha$, $\beta$ and $N^T_i$ enter the second order action without temporal derivatives. So, the
 dynamical degrees of freedom are $\zeta$
    and transverse traceless $h_{ij}$, i.e., scalar and tensor perturbations \cite{Kobayashi:2015gga}:
\begin{subequations}
  \label{jan23-22-5}
  \begin{align}
\label{second_scalar}
\mathcal{S}_{\zeta \zeta}^{(2)}&=\int d\hat{t} d^{3} x  N a^3
\left[
  \frac{\mathcal{G}_S}{N^2}
  \left(\frac{\partial{\zeta}}{\partial \hat{t}}\right)^{2} -
  \frac{\mathcal{F}_S}{a^2} \left (\vec{\nabla} \zeta\right)^{2}\right] \; ,
\\
\mathcal{ S}_{hh}^{(2)}&=\int d\hat{t} d^3x \frac{N a^3}{8}\left[
  \frac{\mathcal{G}_T}{N^2}
            \left (\frac{\partial h_{ij}}{\partial\hat{t}}\right)^2
        -\frac{\mathcal{ F}_T}{a^2}
         h_{ij,k} h_{ij,k}
        \right] \; .
        \label{second_tensor}
  \end{align}
\end{subequations}
We put all the expressions for the coefficients  $\mathcal{G}_S$,  $\mathcal{F}_S$,   $\mathcal{G}_T$,  and
$\mathcal{F}_T$ into Appendix \ref{app:sec_ord}.

\subsection{The model of contracting Universe: properties and constraints}
\label{sec:prop}
In this subsection we briefly list all properties of the contracting Universe solution at early times, which was built and studied in details in Ref.~\cite{Ageeva:2022asq}. Such a solution reads
\begin{align}
    N=\mbox{const} \; , \;\;\;\;\; a= d (-t)^\chi \; , \;\;\;\;\;H=\chi/t,
\label{jan31-22-1}
\end{align}
where $1>\chi > 0$ is a constant and $t= N\hat{t}$ is cosmic time. Note, that since the scale factor $a$ is dimensionless, the dimension of $d$ is as follows: $[d] = \chi$. This kind of early-time cosmology \eqref{jan31-22-1} can be obtained in the framework of the following \textit{Ansatz} for the Lagrangian functions from \eqref{adm_lagr} at $\hat{t}\to -\infty$:
\begin{subequations}
\label{A_old}
	\begin{align}
	&A_2(\hat{t},N) = \hat{g} (-\hat{t}\;)^{-2\mu -2} \cdot a_2 (N) \;\text{,}  \\
	  &A_3 (\hat{t},N)= \hat{g} (-\hat{t}\;)^{-2\mu -1} \cdot a_3 (N)\; \text{,}
          \\
	  A_4 =&A_4 (\;\hat{t}\;)= -B_4(\;\hat{t}\;) = -  \frac{\hat{g}}{2} (-\hat{t}\;)^{-2\mu}\; ,
          \label{A4old}
	\end{align}
\end{subequations}
where $\hat{g}$ is some positive
constant, $\mu$ is another dimensionless parameter of the model, and
\begin{subequations}
  \label{jul18-22-1}
\begin{align}
\label{a_2}
    a_2(N) &= c_2 + \frac{d_2}{N}\; ,\\
    \label{a_3}
    a_3(N) &= c_3+   \frac{d_3}{N} \; ,
\end{align}
\end{subequations}
where $c_2$, $c_3$, $d_2$, $d_3$ are dimensionless constants.
Indeed, it is straightforward to check, that the corresponding equations of motion \eqref{eoms} together with the Ansatz \eqref{A_old}, \eqref{jul18-22-1} admit the solution \eqref{jan31-22-1}. 

Modifying the form of functions \eqref{A_old} and \eqref{jul18-22-1} in a certain way, one can sew the early time contraction \eqref{jan31-22-1} with, for example, the subsequent bounce, expansion stage and etc. Such a construction procedure of the full time evolution can be found in Ref.~\cite{Ageeva:2021yik}.

\subsubsection*{\textit{Stability, no-go theorem and subluminal propagation of perturbations}}
Turning to the second order action for tensor and scalar perturbations around the background solution \eqref{jan31-22-1}, the corresponding coefficients from \eqref{jan23-22-5} are:
\begin{subequations}
\label{jan31-22-3}
\begin{align}
    &\mathcal{G}_T= \mathcal{F}_T =  \frac{g}{(-t)^{2\mu}}\;,
    \label{jan31-22-2}\\
    &\mathcal{G}_S=  g \frac{g_S}{2(-t)^{2\mu}}, \\ 
    &\mathcal{F}_S =  g \frac{f_S}{2(-t)^{2\mu}}\;,
\end{align}
\end{subequations}
where $g = N^{2\mu} \hat{g}$ and $[g] = 2-2\mu$, and the dimensionless coefficients $g_S$ and $f_S$ are
\begin{subequations}
  \label{feb5-22-1}
\begin{align}
  f_S &= \frac{2(2-4 \mu + N^2 a_{3N})}{2\chi - N^2 a_{3N}} = -2\left(\frac{4\mu -2 + d_3}{2\chi  + d_3}\right)\;,
 \label{jan25-22-21a}
  \\
    g_S &= 2   \left[\frac{2 \Big(2 N^3 a_{2N}+ N^4 a_{2NN} - 
    3 \chi (2 \chi + N^3 a_{3NN})\Big)}{(N^2
    a_{3N}-2\chi)^2} + 3\right] \nonumber\\
&= \frac{6 d_3^2}{(2\chi + d_3)^2}\;,
    \label{jan25-22-21b}
\end{align}
\end{subequations}
where again $a_{N} \equiv \partial a/ \partial N$, $a_{NN} \equiv \partial^2 a/ \partial N^2 $, and the initial expressions for these coefficients we put into Appendix \ref{app:sec_ord}; to obtain these $f_S$ and $g_S$ we have used the Ansatz \eqref{A_old}, \eqref{jul18-22-1}.
The corresponding sound speeds of scalar and tensor perturbations are given by
\begin{align}
    \label{us}
 &u_S^2 = \frac{\mathcal{F}_S}{\mathcal{G}_S} = \frac{f_S}{g_S}  ,
\end{align}
\begin{align*}
 &u_T^2 = \frac{\mathcal{F}_T}{\mathcal{G}_T} = 1 .
\end{align*}
So, there is the luminal propagation of the tensor perturbations in our model, while for scalar we stick to the speed smaller than unity, i.e. demand $f_S\leq g_S$ to avoid superluminality.

It is well-known, that the coefficients in second order action are responsible for the stability of the model: one must make sure that $\mathcal{G}_{T,S}>0$, $\mathcal{F}_{T,S}>0$ at each moment of time. Moreover, working in the framework of the Horndeski theory, one should remember about \textit{no-go} argument \cite{Kobayashi:2016xpl}: solutions which involve early non-singular epoch 
suffer from either  singularities or
gradient and/or ghost instabilities at some moment of time in the past or in the future. We go around this problem, imposing \cite{Ageeva:2022asq}
\begin{align*}
    2\mu > \chi +1>0,
\end{align*}
and the detailed discussion of this problem may be found in \cite{Kobayashi:2019hrl,Kobayashi:2016xpl,Libanov:2016kfc,Mironov:2024pjt}.

\subsubsection*{\textit{Power spectrum and amplitudes of primordial perturbations}}
A main advantage of any early Universe model is its success being in agreement with the latest observational data about the inhomogeneity and anisotropy of the cosmic microwave background (CMB) radiation, see some examples of such successful models of inflation and also pre-inflationary phases in \cite{Martin:2013nzq,Kobayashi:2019hrl,Ganz:2022zgs,Frolovsky:2023xid,Choudhury:2024kjj,Ganz:2024ihb} and references therein. As it was shown in Ref.~\cite{Ageeva:2022asq}, the cosmological perturbations with nearly flat power spectrum may be generated at considered early contraction epoch as well.
Indeed, the power spectra for scalar and tensor perturbations in the chosen setup can be written as 
\begin{equation}
  \mathcal{P}_{\zeta} \equiv \mathcal{A}_{\zeta}\left(\frac{k}{k_*}\right)^{n_S-1}\;,
  \;\;\;\;\;  \mathcal{P}_{T} \equiv \mathcal{A}_{T}\left(\frac{k}{k_*}\right)^{n_T} \; ,
    \label{general_ampl}
\end{equation}
where 
where $k_*$ is pivot scale, the spectral tilts are
\begin{equation}
    n_S - 1 = n_T= 2\cdot \left(\frac{1-\mu}{1-\chi}\right) 
    \label{general_n_s}\;,
\end{equation}
Here we would like to emphasize an interesting feature of our model: two tilts are
equal to each other, unlike in most of inflationary models.
The amplitudes are given by \cite{Ageeva:2022asq} (and we also briefly comment the related evaluations in App.~\ref{app:pert})
  \begin{subequations}
    \label{jan25-22-20}
\begin{align}
\label{amplitude}
\mathcal{A}_{\zeta} &= \frac{1}{g}
\cdot\frac{1}{g_S u_{S}^{2 \nu}} \frac{(1-\chi)^{2 \frac{\mu-\chi}{1-\chi}}}{\pi
  \sin ^{2}(\nu \pi) \Gamma^{2}(1-\nu)}\left(\frac{k_*}{2 d}\right)^{2 \frac{1-\mu}{1-\chi}} \; , \\
\mathcal{A}_{T} &= \frac{8}{g} \cdot \frac{(1-\chi)^{2 \frac{\mu-\chi}{1-\chi}}}{\pi
  \sin ^{2}(\nu \pi) \Gamma^{2}(1-\nu)}\left(\frac{k_*}{2 d}\right)^{2 \frac{1-\mu}{1-\chi}},
\label{a_T}
\end{align}
\end{subequations}
where
\begin{align}
    \nu = \frac{1+2 \mu-3 \chi}{2(1-\chi)} = \frac{3}{2}
+ \frac{1-n_S}{2}\; .
\label{feb4-22-1}
\end{align}
We bear in mind, that the corresponding experimental values are \cite{Planck:2018vyg}
\begin{align}
\label{ns}
    n_S = 0.9649 \pm 0.0042,
\end{align}
\begin{align*}
    \mathcal{A}_{\zeta} \simeq 2\cdot 10^{-9},
\end{align*}
i.e. \eqref{general_n_s} together with $\chi>0$ bring that the slightly red tilt can be generated with 
\begin{align}
\label{mu for red tilt}
    \mu > 1 \; .
\end{align}

Another experimentally important value is so called \textit{r-ratio}, which is defined as
\begin{align}
    r = \frac{\mathcal{A}_{T}}{\mathcal{A}_{\zeta}} =
8 \frac{f_S^{\nu}}{g_S^{\nu - 1}}
  =  8 g_S u_S^{2\nu}
\; .
\label{feb1-22-2}
\end{align}
The observations implies \cite{Planck:2018vyg,BICEP:2021xfz,Tristram:2021tvh}
\begin{align}
\label{r}
    r<0.032,
\end{align}
so thinking about the two last equalities in \eqref{feb1-22-2}, one can go two ways:
\begin{enumerate}
    \item Demand $g_S \ll 1$ and $f_S \ll 1$, what is possible to make with the help of much fine-tuning;
    \item Demand $f_S\ll 1$, while $g_S \sim 1$,
and hence $u_S^2\ll 1$. \footnote{Actually, (1) and (2) options together also provide small $r$-ratio.}
\end{enumerate}
In Ref.~\cite{Ageeva:2022asq} we use the second option, what finally leads to a mild fine-tuning
\begin{align}
\label{oct11-24-1}
    d_3 = -2.
\end{align}

It is worth noting here, that small sound speed for scalar perturbation -- for instance, in inflationary models -- usually leads to non-gaussian features, which are incompatible with current observations of the CMB \cite{Planck:2018vyg,BICEP:2021xfz}, see Refs.~\cite{DeFelice:2011zh,Armendariz-Picon:1999hyi,Garriga:1999vw,Seery:2005wm,Chen:2006nt,Peng:2016yvb}. In our model of the contracting Universe \eqref{jan31-22-1} we also stick to the small $u_S$ and in Sec. \ref{sec:eval non gauss} we are going to explore the value of non-Gaussianities to find out if this value is in agreement with observational data  or not. Before that we also provide a brief review on what the non-Gaussianity is in Sec.~\ref{sec:nongauss}.

Finally, let us recall the linearized equation for scalar perturbation \cite{Ageeva:2022asq,Ageeva:2021yik} (also, see the App.\ref{app:pert})
 \begin{equation*}
   \ddot{\zeta}+\frac{2 \mu-3 \chi}{|t|} \dot{\zeta}+\frac{u_{S}^{2} k^{2}}{d^{2} |t|^{2 \chi}} \zeta=0\;,
 \end{equation*}
and here for $0<\chi<1$, the mode $\zeta$ is effectively
 subhorizon at \textit{early times}, i.e.
 the effective Hubble time scale $|t|$ is greater
 than
 the period of oscillations
 $d \cdot |t|^\chi/ (u_S k)$ and the WKB
 approximation can be used here.
Next, at \textit{later times}, the mode is superhorizon and
 we consider the case
\begin{align*}
    2\mu - 3 \chi >0 \; ,
\end{align*}
so that the superhorizon mode
 experiences the Hubble friction and freezes out and related horizon exit time $t_f (k)$ is defined by
 \begin{align*}
      \frac{2\mu - 3\chi}{|t_f|} \sim \frac{u_{S} k}{d |t_f|^{ \chi}},
 \end{align*}
so we emphasize, that with $1> \chi >0$, $2\mu - 3 \chi >0$ 
 the perturbations are
generated such as the weak coupling regime occurs
  all the way down to
  $|t| \sim |t_f|$. In other words, the model does not suffer from the \textit{strong coupling problem} (this problem is under review in the next subsubsection below) at mentioned times.

\subsubsection*{\textit{The legitimacy of the classical treatment}}
The problem of strong coupling manifests itself in considered model when one avoids the no-go theorem. Actually, there are several ways to go around the latter: one \cite{Mironov:2020pqh,Mironov:2024pjt,Cai:2017dyi,Kolevatov:2017voe,Ye:2019sth,Mironov:2019qjt,Ilyas:2020qja,Zhu:2021whu,Mironov:2022ffa,Mironov:2024yqa,Mironov:2024pjj} is to employ beyond Horndeski theories~\cite{Gleyzes:2014dya,Zumalacarregui:2013pma} or 
DHOST generalizations~\cite{Langlois:2015cwa}. Without employing these generalizations, i.e., 
staying within the Horndeski
class like \eqref{Hor_L}, there is still a possibility to avoid the no-go: allow the coefficients
$\mathcal{ G}_T$, $\mathcal{ F}_T$,  $\mathcal{ G}_S$ and  $\mathcal{ F}_S$ to
go to zero as $t\to -\infty$, see Refs.~\cite{Kobayashi:2016xpl, Ageeva:2018lko,Ageeva:2020buc} for details. Since these coefficients serve as the 
analogs of the Planck mass squared, their early-time behavior
$\mathcal{ G}_T, \mathcal{ F}_T,  \mathcal{ G}_S, \mathcal{ F}_S \to 0$ as
$t \to -\infty$
implies
that the \textit{gravitational interaction is strong in the very beginning of the Universe}. 

Bearing in mind, that coefficients $\mathcal{ G}_T, \mathcal{ F}_T,  \mathcal{ G}_S, \mathcal{ F}_S \to 0$ as
$t \to -\infty$, let us note the following fact. In terms of conformal time $\tau  \propto - (-t)^{1-\chi}$ (runs from $-\infty$ to 0, recalling $0<\chi<1$)  the quadratic action
\eqref{second_tensor} for tensor perturbations coincides with the
 General Relativity case
with the background scale factor
\begin{align*}
    a_{E} (\tau) = a(\tau)  \mathcal{ G}_T^{1/2}(\tau)
\propto \frac{1}{(-\tau)^{\frac{\mu - \chi}{1-\chi}}} \; ,
\label{jan24-22-11}
\end{align*}
and this is precisely the scale factor in the Einstein frame in our class of models \eqref{jan31-22-1}. However, depending on $\mu$ the behaviour is different: 
\begin{itemize}
    \item  Indeed, for $\mu <1$ the
  Einstein frame cosmic time is $t_{E} =\int~a_{E}(\tau) d\tau
= - (-\tau)^{\frac{1-\mu}{1-\chi}}$ and it runs
from $t_{E} = - \infty$. Then the effective scale factor increases as
  $a_{E} = (-t_{E})^{-b}$ with $b = \frac{\mu - \chi}{1-\mu} > 1$, see Ref.~\cite{Ageeva:2020gti}. Such a geometry is singular and past geodesically incomplete as
  $t_{E} \to -\infty$. However, the latter is not pathological in
    our case, since by assumption particles with time-independent mass feel
    the Jordan frame geometry rather than the Einstein frame one, see \cite{Rubakov:2022fqk} for details.
    \item Another option is
  $\mu >1$, where one immediately obtains  effective power-law inflation with
  \begin{align*}
       a_{E}(t_E
  ) \propto t_{E}^{\frac{\mu - \chi}{\mu -1}} \; ,
  \end{align*}
  where
  $t_{E} \propto  (-\tau)^{- \frac{\mu -1}{1-\chi}}$ runs from $t_{E} =0$.
  In the Einstein frame, this is a version of
  $G$-inflation~\cite{Kobayashi:2010cm}. In either case,
  for $\mu \approx 1$ the Einstein frame expansion is nearly exponential, so
  the power spectrum of generated tensor perturbations
  is nearly flat. The same thing one can obtain also for the scalar perturbation. 
\end{itemize}

Finally, one can avoid both no-go argument and strong coupling in the ad hoc model with \eqref{A_old}, \eqref{jul18-22-1}. The latter is indeed possible imposing a slow change of the  exponent $\mu = \mu(t)$ from $\mu<1$ in the asymptotic past
   to $\mu > 1$ at later times, when the perturbations are generated, as well as imposing 
\begin{align*}
 0<\chi < \frac{3-n_S}{4- n_S} \approx \frac{2}{3} \; .
\end{align*}
This condition was obtained by the evaluation of the strong
coupling energy scale from the tree-level unitarity bound for tree level matrix elements with cubic order vertices, i.e. using the optical theorem for corresponding partial wave amplitude $\tilde{a}^{(l)}$ \cite{Ageeva:2022asq,Ageeva:2022byg,Grojean:2007zz}:
\begin{align*}
     \mbox{Im}~ \tilde{a}^{(l)} =  \tilde{a}^{(l)}  \tilde{a}^{(l)\, \dagger}, 
\end{align*}
and then comparing found strong coupling energy with the classical scale (which usually related to the Hubble, its time derivative or to the background solution for scalar field $\phi$ and its time derivative as well). The classical treatment is legit when the strong coupling scale is bigger than the classical one. Also, we would like to note here, that it is not necessary consider the higher order interactions terms, since at tree level this terms lead to the weaker condition on the strong coupling absence than the cubic scalar terms, see Refs.~\cite{Ageeva:2020buc, Ageeva:2022asq}. However there are still an open questions about what effect does loops corrections have on the condition of legitimacy of classical description in the case of cosmological models with the strong gravity in the past.

\subsubsection*{\textit{Belinsky--Khalatnikov--Lifshitz phenomenon}}
Here we briefly discuss the BKL phenomenon \cite{Lifshitz:1963ps, Belinsky:1970ew,Belinskii:1972sg} in our setup. Such a problem usually arises in contracting solutions: one of the two
solutions for
a superhorizon mode of given conformal momentum grows as $t$ increases
and diverges in the formal limit $t \to 0$, while  another solution
stays constant in time, what leads to that the Universe becomes strongly
anisotropic and inhomogeneous at late times, which is undesirable
(see, e.g., Ref.~\cite{Erickson:2003zm} for discussion). We can avoid this problem, using
\begin{equation}
\label{BKL_cond}
    2\mu +1 >  3\chi \;,
\end{equation}
for both tensor and scalar perturbations. However, since we already have $\mu >1$ (to obtain the red-tilted spectrum) and $\chi<1$ (to obtain the contracting Universe solution), the condition \eqref{BKL_cond} is automatically satisfied.

\subsection{Covariant Lagrangian of the model}
\label{sec:param}
All properties of contracting Universe, which we discuss in the previous subsection, were entirely studied with the use of ADM Lagrangian \eqref{adm_lagr} with \eqref{A_old}, \eqref{jul18-22-1}. In such setup, we have the following parameters: $\mu$, $\hat{g}$, $c_2$, $d_2$, $c_3$, and $d_3$. However, we have seen that $d_3$ is given by \eqref{oct11-24-1} in order to obtain small scalar sound speed and small $r$-ratio, so for now this parameter is fixed. Also, the smallness of the amplitudes in power spectra of tensor and scalar primordial perturbations fixes the large value of pre-factor $\hat{g}$, see \eqref{jan25-22-20}. 

Now we recall, that actually the Lagrangian in ADM formalism is not unique, i.e. one covariant Lagrangian may lead to several different Lagrangians in ADM form due to, for instance, the different choice of $\hat{t}$ slicing. Thus, to discuss the meaning of $\mu$, $c_2$, $d_2$, and $c_3$ parameters, let us write down the covariant form of Lagrangian \eqref{adm_lagr} with \eqref{A_old} and \eqref{jul18-22-1}:
\begin{subequations}
\label{Horndeski_func_J}
\begin{align}
    &G_2 = \hat{g}\text{e}^{2\mu \phi}\Big\{\text{e}^{ \phi}\left[c_2\text{e}^{ \phi} + (d_2-c_3-2c_3\mu)\sqrt{2X}\right]\nonumber\\
&+2\mu(d_3+2\mu)X\text{ln}X\Big\},\\
    &G_3 = \frac{1}{2}\hat{g}\text{e}^{2\mu \phi}(d_3+2\mu)(2+\text{ln}X),\\
    &G_4 = \frac{1}{2}\hat{g}\text{e}^{2\mu \phi},
\end{align} 
\end{subequations}
which was obtained in \cite{Ageeva:2022asq} with the help of \eqref{jan23-22-1}-\eqref{F}. 
Taking into account the form of the functions \eqref{Horndeski_func_J}, we conclude, that the  \textit{physical} parameters of our model are $c_2$, $\kappa \equiv c_3 (1+2\mu) - d_2$,\footnote{This notation of $\kappa$ was also introduced in Ref.~\cite{Ageeva:2022asq}.} and $\mu$. In the further analysis of the constraints from the non-Gaussianities we will use the convenient combinations of these parameters. For example, for our purposes it will be more convenient to rewrite $c_2$ and $\kappa$ through background parameter $\chi$ \eqref{jan31-22-1}, using equations of motion.  The allowed regions of these parameters can be found imposing all the constraints we have listed in Sec.~\ref{sec:prop}, as well as using the consequence from the equations of motion \eqref{eoms}
with \eqref{oct11-24-1}:
\begin{subequations}
\label{backgroundeq}
\label{chi_N}
  \begin{align}
    \chi &= \frac{3 + 8\rho (\mu-1)(2\mu+1) - \sqrt{9 -
        12\rho (2\mu+1)(5-2\mu)}}{3 + 16\rho (\mu - 1)^2}\; ,
    \\
    N &= \frac{2}{\kappa}\left[1 + 2\mu - 2(\mu -1)\chi\right] \; ,
  \end{align}
\end{subequations}
where $\rho \equiv \frac{c_2}{\kappa^2}$ and we recall that the expression for $N$ is not of particular physical significance (the only requirement is that $N>0$), thus we also will use $N = 1$ everywhere below. Actually, the
initial value of lapse function $N$ could be set to any value by the rescaling of the time.
Thus, without loss of generality one can set this value to be unity. This choice, will not affect any physical results or predictions.
 
In Jordan frame, the standard relation $-\frac{1}{H_{(J)}^2}\frac{dH_{(J)}}{dt_{(J)}}$ which usually relates to first slow-roll parameter in inflation, is bigger than 1 for our model of contracting Universe. However, in Einstein frame, for the case of $\mu>1$ \eqref{mu for red tilt}, the considered model of contracting Universe is power-law inflation, for which $-\frac{1}{H_{(E)}^2}\frac{dH_{(E)}}{dt_{(E)}}$ is much smaller than 1. That is why below we will investigate the current model in Einstein frame and figure out, whether one can use results for non-Gaussianities from the Ref.~\cite{DeFelice:2013ar}, which were obtained under the slow-roll approximation.  To this end, for the case of $\mu>1$ \eqref{mu for red tilt}, we write a covariant form of Lagrangian in Einstein frame as follows:
    \begin{align}
    \label{ya-29-jan-1}
      &\mathcal{S}=\int d^4x \sqrt{-g_{(E)}}
      \Big\{ G^{(E)}_2\left(\phi, X_{(E)}\right)-G^{(E)}_3\left(\phi, X_{(E)}\right)(\Box \phi)^{(E)}\nonumber\\
      &+ G^{(E)}_4 R_{(E)} \Big\}\;,
    \end{align}
where
\begin{subequations}
\label{ya-29-jan-2}
\begin{align}
    &G^{(E)}_2 = \frac{c_{2}e^{-2 (\mu-1) \phi}}{\hat{g}} + \frac{\sqrt{2X_{(E)}} \
d_{2} e^{(1-\mu)\phi} }{ \sqrt{\hat{g}}} \nonumber\\
&-  \frac{\sqrt{2X_{(E)}} c_{3} (1 + 2 \mu) e^{(1-\mu)\phi}  }{ \sqrt{\hat{g}}} - 6 d_{3}{} \mu X_{(E)} - 6 \mu^2 X_{(E)},\\
&G^{(E)}_3 = \frac{1}{2}(d_3 + 2\mu) \text{ln}X_{(E)},\\
&G^{(E)}_4 = \frac{M_{P}^2}{2} = \frac{1}{2},
\end{align}
\end{subequations}
where $M_{P}$ is the reduced Planck mass, which we set equal to 1 in what follows.
The action \eqref{ya-29-jan-1} is obtained with the use of conformal transformation of the form 
    \begin{align*}
        g_{\mu \nu \,  (E)}\; = S^{-2} g_{\mu\nu}, 
    \end{align*}
so
\begin{equation*}
    X_{(E)} = -\frac{1}{2}g_{(E)}^{\mu\nu} \partial_{\mu}\phi\partial_{\nu}\phi=  X S^2,
\end{equation*}
and 
\begin{equation*}
    \sqrt{-g_{(E)}} = S^{-4}\sqrt{-g},
\end{equation*}
with $S = \frac{\text{e}^{-\mu \phi}}{\sqrt{\hat{g}}}$.

The power-law inflation in Einstein frame is a slow-roll inflation model. To illustrate this statement, we follow, for instance, the definitions of Ref.~\cite{DeFelice:2013ar} and consider the corresponding slowly varying parameters:
\begin{subequations}
\label{ya-29-jan-3}
\begin{align}
    &\epsilon \equiv -\frac{1}{H_{(E)}^2}\frac{dH_{(E)}}{dt_{(E)}},\\ 
    &\delta_{\phi} \equiv \frac{1}{H_{(E)}\frac{d\phi}{dt_{(E)}}}\frac{d^2\phi}{dt_{(E)}^2},\\
    &\delta_{G2X} \equiv \frac{X_{(E)}G^{(E)}_{2X_{(E)}}}{H_{(E)}^2 },\\
    &\delta_{G3X} \equiv \frac{\frac{d\phi}{dt_{(E)}}X_{(E)}G^{(E)}_{3X_{(E)}}}{H_{(E)}},
\end{align}
\end{subequations}
where we note that $G^{(E)}_{X_{(E)}} \equiv \partial G^{(E)}/ \partial X_{(E)}$; so instead of standard two slow-roll parameters $\epsilon$ and $\eta$, there are four non-zero small parameters for our inflationary model with the Lagrangian \eqref{ya-29-jan-1}, \eqref{ya-29-jan-2}. The full list of such parameters for general Horndeski theory is presented in Ref.~\cite{DeFelice:2013ar}. Let us derive the explicit formulas for \eqref{ya-29-jan-3} in terms of model parameters.

To that end, we firstly list the corresponding equations of motion in covariant form \cite{Kobayashi:2011nu}
\begin{subequations}
\label{ya-29-jan-4}
\begin{align}
    &2X_{(E)}G^{(E)}_{2X_{(E)}} -G^{(E)}_2\nonumber\\+ &6X_{(E)}\dot{\phi} H_{(E)} G^{(E)}_{3X_{(E)}} - 2X_{(E)}G^{(E)}_{3\phi} -6G^{(E)}_4H_{(E)}^2 =0,
\end{align}
and
\begin{align}
    &G^{(E)}_2 - 2X_{(E)}(G^{(E)}_{3\phi}+\ddot{\phi}G^{(E)}_{3X_{(E)}}) \nonumber\\
    &+ 6G^{(E)}_4H_{(E)}^2+4G^{(E)}_4\dot{H}_{(E)}=0,
\end{align}
\end{subequations}
and where we carefully note, that dot means the derivative with respect to $t_{(E)}>0$ in this Section.
After the substitution of the Lagrangian functions \eqref{ya-29-jan-2}, equations \eqref{ya-29-jan-4} take the form of
\begin{subequations}
\label{ya-29-jan-7}
\begin{align}
    &-\frac{c_2 e^{-2 (\mu -1) \phi
   }}{\hat{g}}-3 \left(H_{(E)}-\mu  \dot{\phi}
   \right) \left(H_{(E)}-(d_3+\mu )
   \dot{\phi} \right) = 0,\\
   &\frac{c_2 e^{-2 (\mu -1) \phi
   }}{\hat{g}}+\frac{\big(d_2-c_3 (2 \mu +1)\big) e^{(1 -\mu)  \phi
   } \dot{\phi}
   }{\sqrt{\hat{g}}}-(d_3+2
   \mu ) \ddot{\phi} \nonumber\\
   &-3 \mu  (d_3+\mu
   ) (\dot{\phi})^2+2 \dot{H}_{(E)}+3 H_{(E)}^2 =0,
\end{align}
\end{subequations}
where we also use $X_{(E)} = \frac{(\dot{\phi})^2}{2}$.
These equations admit the following background solutions:
\begin{subequations}
\label{ya-29-jan-5}
\begin{align}
    &H_{(E)} =  \left(\frac{\mu-\chi}{\mu-1}\right)\frac{1}{t_{(E)}},\\
    &\phi = -\frac{\text{ln}\big[\frac{(\mu-1)t_{(E)}}{\sqrt{\hat{g}}}\big]}{1-\mu}.
\end{align}
\end{subequations}
We emphasize again that $t_{(E)}>0$ here is Einstein frame time.
As a next step, we substitute functions \eqref{ya-29-jan-2},  background solutions \eqref{ya-29-jan-5} as well as \eqref{oct11-24-1} to slowly varying parameters \eqref{ya-29-jan-3} and obtain:
\begin{subequations}
\label{ya-29-jan-6}
\begin{align}
    &\epsilon = \frac{\mu-1}{\mu-\chi},\label{eps thr mu}\\ 
    &\delta_{\phi} = -\epsilon,\\
    &\delta_{G2X} = \frac{d_2-c_3 (2 \mu +1) -6 \mu  (\mu -2)}{2 (\mu -\chi )^2},\label{dg2x}\\
    &\delta_{G3X} = \epsilon \label{g3x thr mu}.
\end{align}
\end{subequations}
These expressions depend on the model parameters, so let us specify the allowed ranges for them. The parameter $\mu$ is related to the experimental range of the scalar spectral index $n_S$ (given in Eq.~\eqref{ns}) through the relation in Eq.~\eqref{general_n_s}. Next, the allowed range for the parameter $\chi$ can be determined by considering the tensor-to-scalar ratio $r$. The latter is constrained by Eq.~\eqref{r}, and in our setup, the $r$-ratio is related to $\chi$ as described in Ref.~\cite{Ageeva:2022asq}:
\begin{align*}
    r = 48 (1-\chi)^{2(\nu-1)} \left( \frac{1-n_S}{3} \right)^\nu,
\end{align*}
where $\nu$ is given by \eqref{feb4-22-1}.
Thus, considering  $r < 0.032 $ we find the lower bound for $\chi$ which is $\chi>0.42$. The upper bound is $ \chi < \frac{3-n_S}{4- n_S} $ and it's evaluation can be found in Ref.~\cite{Ageeva:2022asq}. Summing up, the possible range of $\chi$ for the central value of $n_S = 0.9649$ is
\begin{align*}
    0.42<\chi<0.67.
\end{align*}
The allowed range for $\mu = \frac{1}{2}(3-\chi - n_S + \chi n_S)$ is:
\begin{align*}
    1.0058<\mu<1.01,
\end{align*}
where we also take the central value of $n_S = 0.9649$.
The last step is to simplify the expression for the $\delta_{G2X}$ \eqref{dg2x}. The latter can be done through the use of equations of motion. Indeed, the equations \eqref{ya-29-jan-7} admit the solution \eqref{ya-29-jan-5} in the presence of the following connection between parameters
\begin{align*}
    &3\chi^2 + 3\chi d_3 + c_2 = 0,\\
    &3\chi^2 - (2\mu + 1) (d_3 + 2\chi) - c_3 (1+2\mu) + d_2  + c_2  =0.
\end{align*}
Substituting $d_3 = -2$ \eqref{oct11-24-1} and solving the latter algebraic equations with respect to $c_3$ and $\kappa \equiv c_3 (1+2\mu) - d_2$, we find
\begin{align}
\label{c2 and kappa}
    &c_2 = -3 (\chi ^2-2 \chi ),\\
    &\kappa = 2 +4 \chi -4 \mu  (\chi -1) \nonumber\\
    &=\frac{3 (117 \chi ^2-234 \chi
   +10117)}{5000},
\end{align}
where for $\kappa$ we also use substitution of $\mu = \frac{1}{2}(3-\chi - n_S + \chi n_S)$ and central value $n_S = 0.9649$.
All the parameters \eqref{ya-29-jan-6} are constant, i.e. after all mentioned substitutions of \eqref{c2 and kappa} as well as $\mu = \frac{1}{2}(3-\chi - n_S + \chi n_S)$, $n_S = 0.9649$ the dependence on $\chi$ disappears and we arrive to
\begin{subequations}
\label{numerical small param}
\begin{align}
    &\epsilon =\delta_{G3X}  =\frac{n_\mathcal{S}-1}{n_\mathcal{S}-3} \approx 0.017,\label{eps thr ns}\\
    &\delta_{\phi} = -\epsilon  \approx - 0.017,\\ 
    &\delta_{G2X} = -\frac{3 \left(2500 n_S^2-5000 n_S+2617\right)}{2500
   (n_S-3)^2}\approx -0.035.
\end{align}
\end{subequations}
Thus, we explicitly have shown that all slowly varying  parameters are indeed much smaller than 1 during the Einstein frame power-law inflation. This particularly means that slow-roll approximation as well as results from Ref.~\cite{DeFelice:2013ar} for non-Gaussianity parameter $f_{\text{NL}}$ are applicable for our particular model. For the sake of completeness, in 
Appendix~\ref{app:slowroll}
we also explicitly  show that in Jordan  frame $\epsilon_J = -\frac{\dot{H}_{(J)}}{H_{(J)}^2}$ parameter is greater than unity, i.e in Jordan frame one has contracting Universe for which slow-roll approximation is totally inapplicable. Therefore, one concludes that our particular model does not require a separate numerical investigation of non-Gaussianities. In the next Section we perform analytical evaluations of non-Gaussianity parameters for the chosen model of contracting Universe using the results from Ref.~\cite{DeFelice:2013ar}.

\section{Review on primordial non-Gaussianities}
\label{sec:nongauss}
In this Section we provide a very short review on how to evaluate the deviation from the Gaussian features for primordial perturbation field, what is usually called  \textit{non-Gaussianity} \cite{Komatsu:2001rj,Bartolo:2001cw,Maldacena:2002vr,Creminelli:2003iq,Lyth:2005du,Lyth:2005fi,Byrnes:2006vq,Gangui:1993tt,Verde:1999ij,Creminelli:2005hu,Akama:2019qeh}. 
The amount of non-Gaussianity is defined by the bispectrum of curvature perturbations $\zeta$\footnote{In this paper we consider the scalar primordial perturbations only.}, as
\begin{align}
\label{three_def}
&\braket{\zeta(\vec{k}_{1})\zeta(\vec{k}_{2}) \zeta(\vec{k}_{3})}=(2 \pi)^{3} \delta^{(3)}\left(\vec{k}_{1}+\vec{k}_{2}+\vec{k}_{3}\right) B_{\zeta}\left(k_{1}, k_{2}, k_{3}\right),
\end{align}
where $\zeta(\vec{k})$ is a Fourier component of $\zeta$ with a wave number $\vec{k}$. 
The vacuum expectation value of $\zeta$ for the three-point operator at the conformal time $\tau=\tau_f$ can be expressed as \cite{Seery:2005wm,Chen:2006nt,Maldacena:2002vr}
\begin{align}
\label{exp_val_3}
    &\braket{\zeta(\vec{k}_1)\zeta(\vec{k}_2)\zeta(\vec{k}_3)} \\
&= -i\int^{\tau_f}_{\tau_i}d\tau \;a\;\braket{0|\;[\zeta(\tau_f,\vec{k}_1)\zeta(\tau_f,\vec{k}_2)\zeta(\tau_f,\vec{k}_3),\mathcal{H}_{\text{int}}(\tau)]\;|0},\nonumber
\end{align}
where the Hamiltonian in the interacting picture is given by $\mathcal{H}_{\text{int}} = -\mathcal{L}_{\text{cubic}}$,
and $\tau_i$ is the initial time when the perturbations are deep inside the Hubble radius and $\mathcal{L}_{\text{cubic}}$ is the cubic Lagrangian for perturbations. The latter can be found in Ref.~\cite{DeFelice:2013ar}. 
The bispectrum is given by
\begin{align}
\label{B}
    B_{\zeta}\left(k_{1}, k_{2}, k_{3}\right)=\frac{(2 \pi)^{4}\left(\mathcal{P}_{\zeta}\right)^{2}}{\prod_{i=1}^{3} k_{i}^{3}} \mathcal{F}_{\zeta}\left(k_{1}, k_{2}, k_{3}\right),
\end{align}
where we recall, that $\mathcal{P}_{\zeta}$ is a power spectrum; the size of the three-point function is usually characterized by the number  $f_{\text{NL}}$ (a non-linear parameter), which is given by
\begin{align}
    f_{\text{NL}} = \frac{10}{3}\frac{\mathcal{F}_{\zeta}}{\sum^{3}_{i=1}k_i^3}.
    \label{fnl_def}
\end{align}
The definition of \eqref{fnl_def} is related to \cite{Komatsu:2001rj,Komatsu:2000vy,Komatsu:2001ysk}
\begin{align*}
    \zeta = \zeta_{\text{g}} + \frac{3}{5}f_{\text{NL}} \zeta^2_{\text{g}},
\end{align*}
where $\zeta_{\text{g}}$ is a gaussian field.
The bispectrum can be of different forms or \textit{shapes} depending on the relation between the $\vec{k}_1, \vec{k}_2, \vec{k}_3$. In this paper, we stick to the well-known configurations: local, orthogonal and enfolded. 

The non-linear parameter $f_{\text{NL}}$ is acquired from the three-point correlation function \eqref{three_def}. We follow the same notations and methods as in Ref.~\cite{DeFelice:2013ar} and in the next Section we analytically find the $f_{\text{NL}}$ for different momentum vector configurations. Finally, we apply the constraints on the non-linear parameters from latest observational data \cite{Planck:2018vyg} for our model of contracting Universe \eqref{jan31-22-1} and so derive new bounds on model parameters.

\section{Evaluating non-linear parameters}
\label{sec:eval non gauss}

In this Section we evaluate non-Gaussianity parameters $f_{\text{NL}}$ in our model of contracting Universe \eqref{jan31-22-1}. To this end, we move to the Einstein frame as it was discussed earlier in Sec.~\ref{sec:param} and derive $f_{\text{NL}}$ in the leading order by slowly varying parameters by the same logic as it was done in Ref.~\cite{DeFelice:2013ar}. To find new constraints on model parameters, coming from consideration of non-Gaussianities, we turn to the local, equilateral and enfolded shapes of bispectrum. Each of those are defined by the choice of momentum's configuration, i.e. $k_1=k_2,k_3\rightarrow0$ for local shape, $k_1=k_2=k_3=k$ for equilateral shape and $k_1=k, k_2=k_3=\frac{k}{2}$ for enfolded shape.

\subsubsection*{Local non-Gaussianity}

For local non-Gaussianity one usually expects that the result for single field slow-roll inflationary model will be in agreement with the Maldacena's consistency relation \cite{Maldacena:2002vr}. Let us illustrate that this is indeed the case for our model too.
The leading order (by slowly varying parameters) contribution vanishes for local type, see Ref.~\cite{DeFelice:2013ar}. The next correcting term is \cite{DeFelice:2013ar}
\begin{equation}
\label{fnl local}
    f_{\text{NL}}^{\text{local}}=\frac{5 \epsilon}{6}=\frac{5(\mu-1)}{6(\mu-\chi)} = \frac{5(n_S-1)}{6(n_S-3)},
\end{equation}
where also substitute \eqref{eps thr mu} as well as \eqref{eps thr ns}.
Now recall, that $1 - n_S$ is actually small parameter as well; indeed, for central value $n_S = 0.9649$  one has 
\begin{align}
    1 - n_S \approx 0.035, 
\end{align}
so that $1 - n_S$ smaller than unity. Expanding \eqref{fnl local} with respect to small $1 - n_S$  we  finally arrive to
\begin{equation}
    f_{\text{NL}}^{\text{local}}=\frac{5}{12}(1-n_{S}),
\end{equation}
what is exactly the Maldacena's consistency relation. We also note, that this expression does not depend on the parameters of our model.

\subsubsection*{Equilateral non-Gaussianity}

In the case of equilateral non-Gaussianity the leading by slowly varying parameters order contribution does not vanish. Thus, for the latter we have \cite{DeFelice:2013ar}
\begin{align}
\label{fnl equil}
     f_{\mathrm{NL}}^{\mathrm{equil,lead}}=&\frac{85}{324}\left(1-\frac{1}{u_{S}^{2}}\right)+\frac{20}{81\epsilon_{s}}(1+\lambda_{3X})\delta_{G3X}\nonumber\\
     &+\frac{65}{162u_S^2\epsilon_s}\delta_{G3X},
\end{align}
where $\delta_{G3X}$ is given by \eqref{g3x thr mu} and
\begin{subequations}
\label{for equil}
\begin{align}
        &u_S^2 =\frac{2}{3}(\mu-1)(1-\chi),\\
        &\epsilon_s = \epsilon + \delta_{G3X} = \frac{2(\mu-1)}{\mu-\chi},\\
        &\lambda_{3X}= \frac{X_{(E)}G_{3XX}^{(E)}}{G_{3X}^{(E)}} =-1,
\end{align}
\end{subequations}
and formula for sound speed is obtained from \eqref{feb5-22-1}, \eqref{us} together with \eqref{oct11-24-1}.
Substituting \eqref{g3x thr mu} and \eqref{for equil} into \eqref{fnl equil}, we arrive to
\begin{align}
\label{eq: equil fnl}
    &f_{\mathrm{NL}}^{\mathrm{equil,lead}}=\frac{85}{324}-\frac{5}{81 u_S^2} = \frac{85}{324} + \frac{5}{54 (\mu-1) (\chi -1)}\nonumber\\
    &= \frac{85}{324} + \frac{5}{27 (n_S-1) (\chi -1)^2},
\end{align}
where we also have used $\mu = \frac{1}{2}(3-\chi - n_S + \chi n_S)$ in order to obtain formula in terms of $n_S$.
The observational constraint for $f_{\text{NL}}^{\text{equil}}=-26\pm 47$ \cite{Planck:2018vyg}, so this gives us a new bound on $\mu$ and $\chi$
\begin{equation}
    \mu > 1.0013, \quad \frac{30}{23737 (\mu -1)}+\chi <1,
\end{equation}
where we use \eqref{general_n_s} and apply the following conditions $\mu>1$, $0<\chi<1$.

\subsubsection*{Enfolded non-Gaussianity}

For this type of non-Gaussianity one again has non-zero leading by slowly varying parameters contribution to non-linear $f_{\text{NL}}$ and it reads \cite{DeFelice:2013ar}
\begin{align}
    f_{\mathrm{NL}}^{\text{enfold,lead}}&=\frac{1}{32}\left(1-\frac{1}{u_S^2}\right)+\frac{(1+\lambda_{3X})\delta_{G3X}}{8\epsilon_s},
\end{align}
or, together with \eqref{g3x thr mu} and \eqref{for equil}
\begin{align}
\label{eq: enfold fnl}
     &f_{\mathrm{NL}}^{\text{enfold,lead}}=\frac{1}{32} \left(1-\frac{1}{u_S^2}\right) = \frac{1}{32} \left(1 + \frac{3}{2(\mu-1) (\chi - 1)}\right) \nonumber\\
     &=\frac{1}{32} \left(1 + \frac{3}{(n_S-1) (\chi -1)^2}\right).
\end{align}
The observational constraint for $f_{\text{NL}}^{\text{enfold}}=6\pm30.5$ \cite{Planck:2018vyg} gives us a new condition for parameters  $\mu$ and $\chi$
\begin{equation}
    \mu > 1.0019, \quad \frac{3}{1570 (\mu -1)}+\chi <1,
\end{equation}
where we use \eqref{general_n_s} and apply the following conditions $\mu>1$, $0<\chi<1$.

\subsection{Theoretical and experimental constraints on model parameters}

\begin{figure}[h!]
\centering
\includegraphics[scale=0.45]{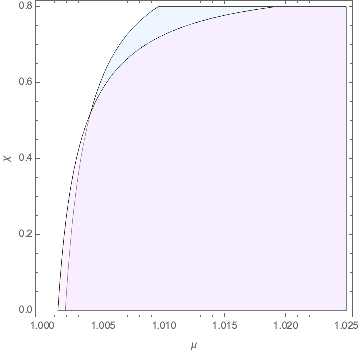}
\caption{Space of parameters $\mu$ and $\chi$ for the Lagrangian \eqref{ya-29-jan-1}, \eqref{ya-29-jan-2}. Blue
 patch is a region where observational constraints for parameters $f_{\text{NL}}^{\text{equil}}$ and $f_{\text{NL}}^{\text{enfold}}$ from \textit{Planck} \cite{Planck:2018vyg} satisfied. Pink patch is a region of $\mu$ and $\chi$ which admits the strong coupling problem absence.}
\label{fig:ng and str}
\end{figure}

\begin{figure}[h!]
\centering
\includegraphics[scale=0.45]{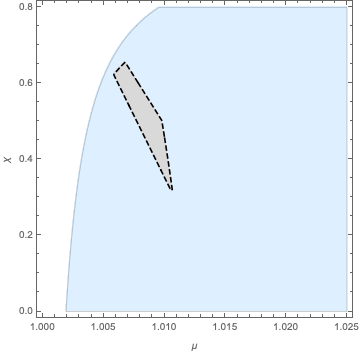}
\caption{Space of parameters $\mu$ and $\chi$ for the Lagrangian \eqref{ya-29-jan-1}, \eqref{ya-29-jan-2}. Blue patch is a region of $\mu$ and $\chi$, where observational constraints for parameters $f_{\text{NL}}^{\text{equil}}$ and $f_{\text{NL}}^{\text{enfold}}$ from \textit{Planck} \cite{Planck:2018vyg} are satisfied. The gray dashed-framed area shows the allowed range of parameters $\mu$ and $\chi$, where all constraints of this Section
are satisfied.}
\label{fig:ng and ok area}
 \end{figure}

\begin{figure}[h!]
\centering
\includegraphics[scale=0.3]{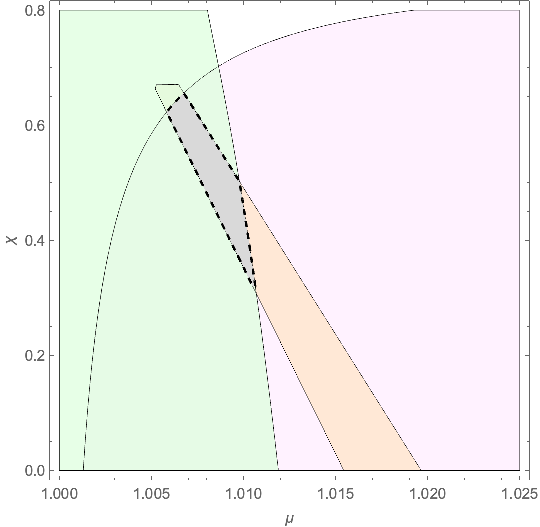}
\caption{Space of parameters $\mu$ and $\chi$ for the Lagrangian \eqref{ya-29-jan-1}, \eqref{ya-29-jan-2}. Pink patch is a region of $\mu$ and $\chi$ which admits the strong coupling problem absence. Green area is allowed parameters which give observationally correct value of $r$-ratio, while orange area relates to the allowed region of parameters where $0<\chi<1$, $2\mu > \chi + 1$, $\mu>1$, $2\mu-3\chi>0$, $2\mu+1>3\chi$, and $0.9607<n_S<0.9691$. The gray dashed-framed area shows the allowed range of parameters $\mu$ and $\chi$, where all constraints of this Section
are satisfied.}
\label{fig:ok area}
\end{figure}

Now, let us once again discuss and list all the constraints (theoretical and observational) we have considered in this Section as well as in Section \ref{sec:prop} and show which of them provide the strongest bound on the model parameters:
\begin{enumerate}
    \item The model of contracting Universe \eqref{jan31-22-1} can be built if one takes
     \begin{equation}
        1>\chi>0.
    \end{equation}
    \item To avoid the no-go theorem from Ref.~\cite{Kobayashi:2016xpl}, one needs to take 
    \begin{equation}
        2\mu>\chi+1>0.
    \end{equation}
    \item The slightly red tilt in spectrum 
can be generated with
\begin{align}
    \mu>1.
\end{align}
    \item The considered superhorizon scalar mode should freeze out at later times where the WKB
approximation can be used; this is ensured if
    \begin{equation}
        2\mu-3\chi>0.
    \end{equation}
    \item Taking into consideration the Belinsky-Khalatnikov-Lifshitz phenomenon \cite{Lifshitz:1963ps,Belinsky:1970ew,Belinskii:1972sg}, which in our case requires that the time-dependent superhorizon solution decays as $t$ runs to zero, we should use
    \begin{equation}
        2\mu+1>3\chi.
    \end{equation}
    \item The requirement for $r$-ratio to satisfy the observational constraint \cite{Tristram:2021tvh} reads
    \begin{equation}
    \label{final r}
        r<0.032.
    \end{equation}
For our specific model the expression for $r$-ratio in terms of model parameters is \cite{Ageeva:2022asq}:
\begin{align}
    r=48(1-\chi)^{2(\nu-1)}\left(\frac{1-n_S}{3}\right)^\nu,
\end{align}
where $\nu=2-\frac{n_\mathrm{S}}{2}$.
    \item The condition of the strong-coupling absence reads \cite{Ageeva:2022asq}
    \begin{align}
    \label{final no SC}
         \frac{E_{\mathrm{strong}}}{E_{cl}}=\tilde{C}\cdot\left(\frac{r^{4/\nu}}{\mathcal{A}_\zeta}\right)^{1/6}>1,
    \end{align}
where
   \begin{widetext}
   \begin{align}
            &\tilde{C}=2^{\frac{12-11n_S}{24-6n_S}}3^{\frac{4-3n_S}{24-6n_S}}\left(1-\chi\right)^{\frac{12-n_S}{3(4-n_S)}}\left(\frac{(2-2\chi)^{1-n_S}\big[2+(1-n_S)-\chi\big(3+(1-n_\mathcal{S})\big)\big]^{-(1-n_\mathcal{S})}}{\Gamma^2(\frac{n_\mathcal{S}}{2}-1)\mathrm{sin}^2\left[(2-\frac{n_\mathcal{S}}{2})\pi\right]}\right)^{1/6}.
      \end{align}
        \end{widetext}
    \item Finally, in this Section we have obtained two new bounds from equilateral and enfolded non-Gaussianities, which read
\begin{subequations}
\label{final nongauss}
\begin{align}
      &\mu > 1.0013, \quad \frac{30}{23737 (\mu -1)}+\chi <1,\\
      &\mu > 1.0019, \quad \frac{3}{1570 (\mu -1)}+\chi <1,
\end{align}
\end{subequations}
The second inequality corresponds enfolded non-Gaussianities and it gives the stronger bounds on model parameters in comparison with the equilateral one.
The local non-Gaussianities do not provide any bounds on parameters, since in the leading non-zero order the Maldacena's consistency relation holds. 
\end{enumerate}
We show, how these listed constrains are related to each other in Figs.~\ref{fig:ng and str} - \ref{fig:ok area}. The Fig.~\ref{fig:ng and str} includes the comparison between bounds on $\mu$ and $\chi$ from strong coupling absence \eqref{final no SC} and non-Gaussianities \eqref{final nongauss}. We note that, marginally, both constraints are the same, which is not surprising. Indeed, small non-Gaussianities imply that the cubic interaction terms are sufficiently small. Similarly, the unitarity bounds effectively convey the same idea: the smaller the interaction terms, the less likely they are to break unitarity. The Fig.~\ref{fig:ng and ok area} gives the blue allowed region of $\mu$ and $\chi$ parameters from non-Gaussianities \eqref{final nongauss} together with the gray patch where $\mu$ and $\chi$ respect all the constraints we have listed above. Finally,  Fig.~\ref{fig:ok area} includes the intersection of patches from strong coupling absence \eqref{final no SC} (pink one), observationally allowed $r$-ratio \eqref{final r} (green one) and again gray patch where $\mu$ and $\chi$ respect all the listed above constraints. From this Figure \ref{fig:ok area} we see, that the observational constraints for $n_{S}$ and $r$-ratio \textit{ensure that bounds from non-Gaussianities hold}. We conclude, that one can take $\mu$ and $\chi$ parameters from healthy gray region (Fig.~\ref{fig:ng and ok area} and Fig.~\ref{fig:ok area}) and then obtain stable contracting Universe, where all (today available) observational bounds are respected while the classical description is applicable as well.

\section{Conclusion}
\label{sec:summary}
In this paper, we have studied the non-Gaussianities in a novel non-singular cosmological scenario proposed in Ref.~\cite{Ageeva:2022asq}. This particular model is of considerable interest for several reasons. First, it offers a viable alternative to inflation: a contracting Universe followed by a bounce. The scenario is formulated within the framework of pure Horndeski gravity, involving the functions $G_2(\phi,X)$, $G_3(\phi,X)$, and $G_4(\phi)$, which depend solely on the scalar field $\phi$. In this sense, the model is as minimal as possible, avoiding extensions beyond Horndeski gravity or the introduction of DHOST theories.

Second, the scenario yields predictions consistent with current observational data. It provides acceptable values for the tensor-to-scalar ratio $r$, the scalar spectral index $n_S$, and the scalar amplitude $\mathcal{A}_{\zeta}$, thus aligning well with results from Planck and BICEP/Keck experiments ~\cite{Planck:2018vyg,BICEP:2021xfz}.

In this setup, the small tensor-to-scalar ratio is achieved at the expense of a small (in comparison with unity) sound speed for scalar perturbations. However, as seen from Eqs.~\eqref{eq: equil fnl} and \eqref{eq: enfold fnl}, a small sound speed typically enhances non-Gaussianities. Therefore, it is essential to verify whether the model remains consistent with experimental bounds on non-Gaussianity. We carry out this analysis and find that observational constraints on $f_{\text{NL}}$ indeed impose non-trivial bounds on the model parameters.

In addition to experimental bounds on non-Gaussianities, there are also purely theoretical constraints on the model parameters. In particular, the Bounce Universe scenario can be refered as the scenario with the strong gravity in the past. While this feature allows the model to circumvent the no-go theorem, it simultaneously introduces the risk of entering a strong-coupling regime at early times.

To maintain unitarity in the asymptotic past, one must impose quite restrictive conditions on the model parameters. Interestingly enough, we find that both the unitarity bounds and the observational bounds on non-Gaussianities yield constraints of the same form, see Fig.~\ref{fig:ng and str}. There we denote a region where observational constraints for
non-linear parameters are satisfied as a blue patch and 
a region of model parameters which admit the strong coupling problem
absence as a pink area. This correspondence is not coincidental. The smallness of non-Gaussianities implies that the cubic interaction terms are suppressed. At the same time, the unitarity bounds, which follow from the optical theorem, require that the matrix elements remain sufficiently small—again implying that interaction terms must be sufficiently small.

However, it is important to emphasize that these two constraints stem from fundamentally different sources: the unitarity bounds are theoretical, while the bounds on non-Gaussianities are derived from observational data. As a result, although they may take similar forms, they do not necessarily coincide. In this paper, we explicitly observe such a scenario; see Fig.~\ref{fig:ng and str}.

After evaluating all the constraints, we find that the bounds on the scalar spectrum tilt, the tensor-to-scalar ratio, and the unitarity bounds together ensure that the observational constraints on non-Gaussianities are trivially satisfied, as shown in Figs.~\ref{fig:ng and ok area} and~\ref{fig:ok area}. Thus, we have explicitly verified that the current model satisfies modern observational constraints on primordial non-Gaussianities. Therefore, the Bounce Universe scenario with strong gravity in the past from Ref.~\cite{Ageeva:2022asq} is fully viable and is consistent with current observational data, including the scalar spectrum tilt, the amplitude of scalar perturbations, the tensor-to-scalar ratio, and scalar non-Gaussianities.

\section*{Acknowledgments} 

The authors are grateful to Shingo Akama, Sergei Mironov and Victoria Volkova for useful comments and fruitful discussions. The work of PP on Sec. 1 has been supported by IBS under the project code IBS-R018-D3. The work of YA and MK on Sec. 2 and Sec. 3 has been supported by Russian Science
Foundation Grant No. 24-72-00121, https://rscf.ru/project/24-72-00121/. The work of MK on Sec. 4 has been supported by Theoretical
Physics and Mathematics Advancement Foundation ``BASIS''.

\appendix

\section{General expressions for the second order action coefficients}
\label{app:sec_ord}
\numberwithin{equation}{section}

\setcounter{equation}{0}
The purpose of this Appendix is to list the general formulas for the coefficients in \eqref{jan23-22-5}. 
These are in ADM formalism as well as under the assumption, that $A_4$ and $B_4$ do not depend on $N$: \footnote{More general formulas can be found, for example, in Ref.~\cite{Kobayashi:2015gga}.}
\begin{subequations}
\begin{align*}
         \mathcal{ G}_T &=  -2A_4\;,\\
         \mathcal{ F}_T &= 2B_4\;,
\end{align*}
\end{subequations}
 and
\begin{subequations}
    \begin{eqnarray*}
        \mathcal{ F}_S&=&\frac{1}{a N}\frac{d}{d \hat{t}}\left(\frac{a}{\Theta}\mathcal{ G}_T^2\right)
        -\mathcal{ F}_T\;, 
        \label{eq:Fs_form}
        \\
        \mathcal{ G}_S&=&\frac{\Sigma }{\Theta^2}\mathcal{ G}_T^2+3\mathcal{ G}_T\;, \label{eq:Gs_form}
    \end{eqnarray*}
    \end{subequations}
 with
\begin{subequations}
 \begin{align*}
      \Sigma&=
      N A_{2N}+\frac{1}{2}N^2A_{2NN}+
      \frac{3}{2}N^2A_{3NN}H+6A_4H^2\;, 
        \\
        \Theta&=2H\Big(\frac{NA_{3N}}{4H}-A_4 \Big) .
        \label{theta}
    \end{align*}
\end{subequations}
Substitution of the concrete Ansatz \eqref{A_old} and \eqref{jul18-22-1} leads to \eqref{jan31-22-2}-\eqref{jan25-22-21b}, while using \eqref{oct11-24-1} immediately leads to simpler formulas for coefficients $f_S$ and $g_S$ \cite{Ageeva:2022asq}:
\begin{subequations}
\label{fs-feb6}
  \begin{align}
    f_S &= \frac{4(\mu -1)}{1-\chi} = 2(1 - n_S)\; ,\\
    g_S &= \frac{6 }{(1 -\chi)^2} \; ,
\end{align}
\end{subequations}
which we have used for the power spectrum and bispectrum calculations.

\section{Generating perturbations: linearized equation and its solution}
\label{app:pert}
\numberwithin{equation}{section}

\setcounter{equation}{0}
This Appendix includes the explicit formulas for the solution of linearized equation for primordial perturbation $\zeta$. Almost the same formulas were obtained in \cite{Ageeva:2022asq}. To this end, we follow the textbook evaluations
(see Ref.~\cite{Gorbunov:2011zzc} for more details) and consider:
\begin{equation}
\label{eq_zeta}
    \frac d{dt}(a^3\mathcal{G}_S\dot\zeta)-a\mathcal{G}_Su_s^2\partial^2\zeta=0,
\end{equation}
which leads from \eqref{second_scalar} (and turning to cosmic time).
After the Fourier transform one arrives to
\begin{align}
\label{zeta_fourier}
    &\zeta(\tau,\vec{x})=\frac{1}{(2\pi)^3}\int d^3\vec{k}\zeta(\tau,\vec{k})e^{i\vec{k}\cdot\vec{x}},\\
&\zeta(\tau,\vec{k})=u(\tau,\vec{k})a(\vec{k})+u^*(\tau,-\vec{k})a^\dagger(-\vec{k}),\nonumber
\end{align}
where the ladder operators $a(\vec{k})$ and  $a^\dagger(-\vec{k})$ are defined as
\begin{align*}
    &\left[a(k_1),a^\dagger(k_2)\right] =(2\pi)^3\delta^{(3)}(k_1-k_2),\\
    &[a(k_1),a(k_2)]=\left[a^\dagger(k_1),a^\dagger(k_2)\right]=0.
\end{align*}
The equation \eqref{eq_zeta} can be written for $u$
\begin{equation*}
\ddot{u}+\frac{(a^3\mathcal{G}_S)^\cdot}{a^3\mathcal{G}_S}\dot{u}+u_s^2\frac{k^2}{a^2}u=0,
\end{equation*}
and then in conformal time $\tau=\int a^{-1}dt$ with the change $v=uz$, $z=a\sqrt{2\mathcal{G}_S}$ it is
\begin{equation}
\label{ur v s z}
    v''+\left(u_s^2k^2-\frac{z''}{z}\right)v=0,
\end{equation}
where prime means the derivative with respect to conformal time; also
\begin{align*}
        &z=d\sqrt{gg_s}\big[d(1-\chi)(-\tau)\big]^{\frac{\chi-\mu}{1-\chi}},\\
        &\frac{z''}{z}=\frac{(n_S-3)(n_S-5)}{4 \tau^2},
\end{align*}
where we recall that $n_S=0.9649\pm0.0042$ and $n_S-1=n_T=2\cdot\left(\frac{1-\mu}{1-\chi}\right)$.
Performing another change of variables $v=f\sqrt{-\tau}$, rewrite once again an equation \eqref{ur v s z} and arrive to
\begin{equation*}
\begin{aligned}
    &x^2f_{xx}+xf_{x}+(x^2-\nu^2)f=0, \quad x = -u_sk\tau,\quad
    \nu^2=\frac{(n_S - 4)^2}{4},
    \end{aligned}
\end{equation*}
what is the Bessel equation and the solution is the Hankel function of the first kind (we take into account the first kind only in order to obtain the right asymptotic behaviour, which we discuss shortly below)
\begin{align}
\label{hankel}
        &u=\frac{C\sqrt{-\tau}}{a \sqrt{2\mathcal{G_S}}}H_{\nu}^{(1)}(-u_sk\tau)\nonumber\\
&=C\sqrt{-\tau}\frac{\big[d(1-\chi)(-\tau)\big]^{\frac{3-n_S}{2}}}{d\sqrt{gg_s}}H_{\nu}^{(1)}(-u_sk\tau),
\end{align}
where we also turning back to the function $u$.
To find the constant $C$ in this solution, we consider the asymptotic\\
$k\tau\to-\infty $, so \eqref{ur v s z}  is nothing but a wave equation
with a solution 
\begin{equation*}
\begin{aligned}
   & v''+u_S^2k^2v=0,\\
    &v_{wave}(x)=\int\frac{d^3k}{(2\pi)^3}\frac{1}{\sqrt{2E_k}}(a_{\vec{k}}e^{-iu_skx}+a_{\vec{k}}^*e^{iu_skx}),\\
   & v_{wave}(k,\tau)\Big|_{\tau\to-\infty}=\frac{e^{-iu_Sk\tau}}{\sqrt{2u_sk}}.
    \end{aligned}
\end{equation*}
After some straightforward evaluations we arrive to
\begin{align*}
   &C=\frac{\sqrt{\pi}}{2}e^{i\frac{2\nu+1}{4}\pi}.
\end{align*}
We finish this Appendix with the late time asymptotics 
(formally, $|\tau| \to 0$), where the solution \eqref{hankel} becomes
time-independent
\begin{equation*}
  \zeta=
  (- i) \frac{\mathfrak{C}}{\sin (\nu \pi)} \frac{(1-\chi)^{\nu}}{u_{S}^{\nu} \Gamma(1
  -\nu)}\left(\frac{2 d}{k}\right)^{\nu},
\end{equation*}
with
\begin{equation*}
  \mathfrak{C}=\frac{1}{(g g_S)^{1 / 2}}
  \frac{1}{2^{5 / 2} \pi(1-\chi)^{1 / 2}} \frac{1}{d^{3 / 2}}\;,
\end{equation*}
up to an irrelevant time-independent phase.
Through the formula
\begin{equation*}
\mathcal{P}_{\zeta}=4 \pi k^{3} \zeta^{2}\;,
\end{equation*}
collecting all factors one obtains the result
\eqref{general_ampl}.

\section{Inapplicability of slow-roll
approximations in Jordan frame}
\label{app:slowroll}
\numberwithin{equation}{section}

\setcounter{equation}{0}

In this Appendix we discuss important feature of contracting Universe in Jordan frame. Precisely, we will show, that in this frame it is impossible to  compute the scalar bispectrum by taking into account slow-roll
corrections as it was done in Ref.~\cite{DeFelice:2013ar}. 

\begin{figure}
\centering
\includegraphics[width=0.4\textwidth]{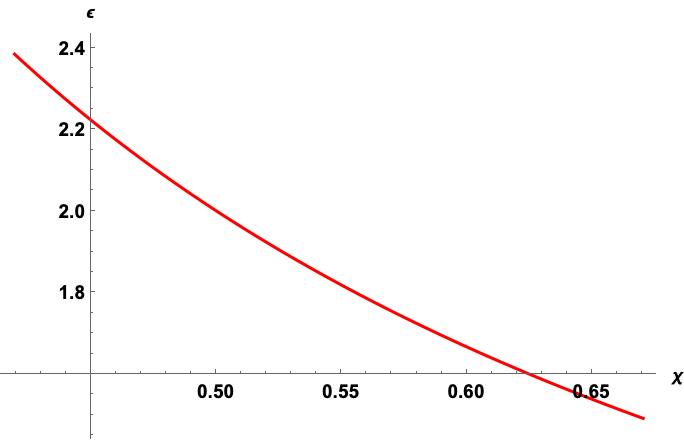}
\caption{The dependence on $\chi$ for $\epsilon = - \frac{\dot{H}}{H^2}$ parameter: Jordan frame.}
\label{fig:eps}
\end{figure}

To that end, we demonstrate that the discussed in Section \ref{sec:param} relation $- \frac{\dot{H}}{H^2}$ is bigger than one in Jordan frame as we expect: the Universe passes through the contraction and the Hubble parameter changes fast with time. The Lagrangian functions in this frame is given by \eqref{Horndeski_func_J}. We recall, that corresponding solution of equations of motion is
\begin{align*}
    &H =  \frac{\chi}{t},\\
    &\phi = -\text{ln}(-t),
\end{align*}
and in this subsection all values are related to Jordan frame, so time $t$ is a cosmic time in Jordan frame (and we recall that we work with $N=1$).
Proceeding the same steps as for Einstein frame, we find out the behavior of parameter $\epsilon \equiv - \frac{\dot{H}}{H^2}$ as a function of $\chi$, see Fig.~\ref{fig:eps}, and prove that this relation is always bigger than 1 within the permitted range of $\chi$. 


\begin{thebibliography}{99}
\bibitem{Starobinsky:1980te}
A.~A.~Starobinsky,
Phys. Lett. B \textbf{91}, 99-102 (1980)
doi:10.1016/0370-2693(80)90670-X

\bibitem{Guth:1980zm}
A.~H.~Guth,
Phys. Rev. D \textbf{23}, 347-356 (1981)
doi:10.1103/PhysRevD.23.347

\bibitem{Sato:1981qmu}
K.~Sato,
Mon. Not. Roy. Astron. Soc. \textbf{195}, no.3, 467-479 (1981)
doi:10.1093/mnras/195.3.467

\bibitem{Linde:1981mu}
A.~D.~Linde,
Phys. Lett. B \textbf{108}, 389-393 (1982)
doi:10.1016/0370-2693(82)91219-9

\bibitem{Albrecht:1982wi}
A.~Albrecht and P.~J.~Steinhardt,
Phys. Rev. Lett. \textbf{48}, 1220-1223 (1982)
doi:10.1103/PhysRevLett.48.1220

\bibitem{Planck:2018vyg}
N.~Aghanim \textit{et al.} [Planck],
Astron. Astrophys. \textbf{641}, A6 (2020)
[erratum: Astron. Astrophys. \textbf{652}, C4 (2021)]
doi:10.1051/0004-6361/201833910
[arXiv:1807.06209 [astro-ph.CO]].

\bibitem{BICEP:2021xfz}
P.~A.~R.~Ade \textit{et al.} [BICEP and Keck],
Phys. Rev. Lett. \textbf{127}, no.15, 151301 (2021)
doi:10.1103/PhysRevLett.127.151301
[arXiv:2110.00483 [astro-ph.CO]].

\bibitem{Penrose:1964wq}
R.~Penrose,
Phys. Rev. Lett. \textbf{14}, 57-59 (1965)
doi:10.1103/PhysRevLett.14.57

\bibitem{Hawking:1966vg}
S.~W.~Hawking,
Phys. Rev. Lett. \textbf{17}, 444-445 (1966)
doi:10.1103/PhysRevLett.17.444

\bibitem{Borde:1993xh}
A.~Borde and A.~Vilenkin,
Phys. Rev. Lett. \textbf{72}, 3305-3309 (1994)
doi:10.1103/PhysRevLett.72.3305
[arXiv:gr-qc/9312022 [gr-qc]].

\bibitem{Borde:1996pt}
A.~Borde and A.~Vilenkin,
Int. J. Mod. Phys. D \textbf{5}, 813-824 (1996)
doi:10.1142/S0218271896000497
[arXiv:gr-qc/9612036 [gr-qc]].

\bibitem{Borde:2001nh}
A.~Borde, A.~H.~Guth and A.~Vilenkin,
Phys. Rev. Lett. \textbf{90}, 151301 (2003)
doi:10.1103/PhysRevLett.90.151301
[arXiv:gr-qc/0110012 [gr-qc]].

\bibitem{Yoshida:2018ndv}
D.~Yoshida and J.~Quintin,
Class. Quant. Grav. \textbf{35}, no.15, 155019 (2018)
doi:10.1088/1361-6382/aacf4b
[arXiv:1803.07085 [gr-qc]].

\bibitem{Arefeva:2011qvf}
I.~Y.~Aref'eva and I.~V.~Volovich,
JHEP \textbf{08}, 102 (2011)
doi:10.1007/JHEP08(2011)102
[arXiv:1103.0273 [hep-th]].

\bibitem{Arefeva:2011yua}
I.~Y.~Aref'eva, N.~V.~Bulatov and R.~V.~Gorbachev,
Theor. Math. Phys. \textbf{173}, 1466-1480 (2012)
doi:10.1007/s11232-012-0127-1
[arXiv:1112.5951 [hep-th]].

\bibitem{Silva:2015qna}
C.~A.~S.~Silva,
Eur. Phys. J. C \textbf{78}, no.5, 409 (2018)
doi:10.1140/epjc/s10052-018-5882-1
[arXiv:1503.00559 [gr-qc]].

\bibitem{Silva:2020bnn}
C.~Silva,
Phys. Rev. D \textbf{102}, no.4, 046001 (2020)
doi:10.1103/PhysRevD.102.046001
[arXiv:2008.07279 [gr-qc]].

\bibitem{Silva:2023ieb}
C.~Silva,
Nucl. Phys. B \textbf{998}, 116402 (2024)
doi:10.1016/j.nuclphysb.2023.116402
[arXiv:2312.05260 [gr-qc]].

\bibitem{Hawking:1973uf}
S.~W.~Hawking and G.~F.~R.~Ellis,
Cambridge University Press, 2023,
ISBN 978-1-009-25316-1, 978-1-009-25315-4, 978-0-521-20016-5, 978-0-521-09906-6, 978-0-511-82630-6, 978-0-521-09906-6
doi:10.1017/9781009253161

\bibitem{Tipler:1978zz}
F.~J.~Tipler,
Phys. Rev. D \textbf{17}, 2521-2528 (1978)
doi:10.1103/PhysRevD.17.2521

\bibitem{Rubakov:2014jja}
V.~A.~Rubakov,
Phys. Usp. \textbf{57}, 128-142 (2014)
doi:10.3367/UFNe.0184.201402b.0137
[arXiv:1401.4024 [hep-th]].

\bibitem{Horndeski:1974wa}
G.~W.~Horndeski,
Int. J. Theor. Phys. \textbf{10}, 363-384 (1974)
doi:10.1007/BF01807638

\bibitem{Kobayashi:2019hrl}
T.~Kobayashi,
Rept. Prog. Phys. \textbf{82}, no.8, 086901 (2019)
doi:10.1088/1361-6633/ab2429
[arXiv:1901.07183 [gr-qc]].

\bibitem{Battefeld:2014uga}
D.~Battefeld and P.~Peter,
Phys. Rept. \textbf{571}, 1-66 (2015)
doi:10.1016/j.physrep.2014.12.004
[arXiv:1406.2790 [astro-ph.CO]].

\bibitem{Lehners:2008vx}
J.~L.~Lehners,
Phys. Rept. \textbf{465}, 223-263 (2008)
doi:10.1016/j.physrep.2008.06.001
[arXiv:0806.1245 [astro-ph]].

\bibitem{Brandenberger:2016vhg}
R.~Brandenberger and P.~Peter,
Found. Phys. \textbf{47}, no.6, 797-850 (2017)
doi:10.1007/s10701-016-0057-0
[arXiv:1603.05834 [hep-th]].

\bibitem{Akama:2025ows}
S.~Akama,
JCAP \textbf{06}, 063 (2025)
doi:10.1088/1475-7516/2025/06/063
[arXiv:2502.14850 [astro-ph.CO]].

\bibitem{Creminelli:2010ba}
P.~Creminelli, A.~Nicolis and E.~Trincherini,
JCAP \textbf{11}, 021 (2010)
doi:10.1088/1475-7516/2010/11/021
[arXiv:1007.0027 [hep-th]].

\bibitem{Creminelli:2012my}
P.~Creminelli, K.~Hinterbichler, J.~Khoury, A.~Nicolis and E.~Trincherini,
JHEP \textbf{02}, 006 (2013)
doi:10.1007/JHEP02(2013)006
[arXiv:1209.3768 [hep-th]].

\bibitem{Nishi:2015pta}
S.~Nishi and T.~Kobayashi,
JCAP \textbf{03}, 057 (2015)
doi:10.1088/1475-7516/2015/03/057
[arXiv:1501.02553 [hep-th]].

\bibitem{GilChoi:2025hbs}
H.~Gil Choi, P.~Petrov and M.~Yamaguchi,
JHEP \textbf{08} (2025), 044
doi:10.1007/JHEP08(2025)044
[arXiv:2503.02626 [hep-th]].

\bibitem{Ostrogradsky:1850fid}
M.~Ostrogradsky,
Mem. Acad. St. Petersbourg \textbf{6}, no.4, 385-517 (1850)

\bibitem{Kobayashi:2016xpl}
T.~Kobayashi,
Phys. Rev. D \textbf{94}, no.4, 043511 (2016)
doi:10.1103/PhysRevD.94.043511
[arXiv:1606.05831 [hep-th]].

\bibitem{Libanov:2016kfc}
M.~Libanov, S.~Mironov and V.~Rubakov,
JCAP \textbf{08}, 037 (2016)
doi:10.1088/1475-7516/2016/08/037
[arXiv:1605.05992 [hep-th]].

\bibitem{Gleyzes:2014dya}
J.~Gleyzes, D.~Langlois, F.~Piazza and F.~Vernizzi,
Phys. Rev. Lett. \textbf{114}, no.21, 211101 (2015)
doi:10.1103/PhysRevLett.114.211101
[arXiv:1404.6495 [hep-th]].

\bibitem{Langlois:2015cwa}
D.~Langlois and K.~Noui,
JCAP \textbf{02}, 034 (2016)
doi:10.1088/1475-7516/2016/02/034
[arXiv:1510.06930 [gr-qc]].

\bibitem{Kolevatov:2017voe}
R.~Kolevatov, S.~Mironov, N.~Sukhov and V.~Volkova,
JCAP \textbf{08}, 038 (2017)
doi:10.1088/1475-7516/2017/08/038
[arXiv:1705.06626 [hep-th]].

\bibitem{Mironov:2020pqh}
S.~Mironov, V.~Rubakov and V.~Volkova,
JHEP \textbf{04}, 035 (2021)
doi:10.1007/JHEP04(2021)035
[arXiv:2011.14912 [hep-th]].

\bibitem{Mironov:2022ffa}
S.~Mironov and V.~Volkova,
Int. J. Mod. Phys. A \textbf{37}, no.14, 2250088 (2022)
doi:10.1142/S0217751X22500889
[arXiv:2204.05889 [hep-th]].

\bibitem{Mironov:2024yqa}
S.~Mironov, M.~Sharov and V.~Volkova,
Eur. Phys. J. C \textbf{85}, no.1, 50 (2025)
doi:10.1140/epjc/s10052-024-13730-5
[arXiv:2408.01480 [gr-qc]].

\bibitem{Mironov:2024pjt}
V.~E.~Volkova and S.~A.~Mironov,
Usp. Fiz. Nauk \textbf{195}, no.2, 163-176 (2025)
doi:10.3367/UFNr.2024.12.039826
[arXiv:2409.16108 [gr-qc]].

\bibitem{Ageeva:2022asq}
Y.~Ageeva, P.~Petrov and V.~Rubakov,
JHEP \textbf{01}, 026 (2023)
doi:10.1007/JHEP01(2023)026
[arXiv:2207.04071 [hep-th]].

\bibitem{Meerburg:2009ys}
P.~D.~Meerburg, J.~P.~van der Schaar and P.~S.~Corasaniti,
JCAP \textbf{05}, 018 (2009)
doi:10.1088/1475-7516/2009/05/018
[arXiv:0901.4044 [hep-th]].

\bibitem{DeFelice:2011zh}
A.~De Felice and S.~Tsujikawa,
JCAP \textbf{04}, 029 (2011)
doi:10.1088/1475-7516/2011/04/029
[arXiv:1103.1172 [astro-ph.CO]].

\bibitem{DeFelice:2013ar}
A.~De Felice and S.~Tsujikawa,
JCAP \textbf{03}, 030 (2013)
doi:10.1088/1475-7516/2013/03/030
[arXiv:1301.5721 [hep-th]].

\bibitem{Maldacena:2002vr}
J.~M.~Maldacena,
JHEP \textbf{05}, 013 (2003)
doi:10.1088/1126-6708/2003/05/013
[arXiv:astro-ph/0210603 [astro-ph]].

\bibitem{Chowdhury:2015cma}
D.~Chowdhury, V.~Sreenath and L.~Sriramkumar,
JCAP \textbf{11} (2015), 002
doi:10.1088/1475-7516/2015/11/002
[arXiv:1506.06475 [astro-ph.CO]].

\bibitem{Nandi:2019xag}
D.~Nandi and L.~Sriramkumar,
Phys. Rev. D \textbf{101} (2020) no.4, 043506
doi:10.1103/PhysRevD.101.043506
[arXiv:1904.13254 [gr-qc]].

\bibitem{Creminelli:2004yq}
P.~Creminelli and M.~Zaldarriaga,
JCAP \textbf{10}, 006 (2004)
doi:10.1088/1475-7516/2004/10/006
[arXiv:astro-ph/0407059 [astro-ph]].

\bibitem{Seery:2005wm}
D.~Seery and J.~E.~Lidsey,
JCAP \textbf{06}, 003 (2005)
doi:10.1088/1475-7516/2005/06/003
[arXiv:astro-ph/0503692 [astro-ph]].

\bibitem{Chen:2006nt}
X.~Chen, M.~x.~Huang, S.~Kachru and G.~Shiu,
JCAP \textbf{01}, 002 (2007)
doi:10.1088/1475-7516/2007/01/002
[arXiv:hep-th/0605045 [hep-th]].

\bibitem{Cheung:2007sv}
C.~Cheung, A.~L.~Fitzpatrick, J.~Kaplan and L.~Senatore,
JCAP \textbf{02}, 021 (2008)
doi:10.1088/1475-7516/2008/02/021
[arXiv:0709.0295 [hep-th]].

\bibitem{Ganc:2010ff}
J.~Ganc and E.~Komatsu,
JCAP \textbf{12}, 009 (2010)
doi:10.1088/1475-7516/2010/12/009
[arXiv:1006.5457 [astro-ph.CO]].

\bibitem{Renaux-Petel:2010paw}
S.~Renaux-Petel,
JCAP \textbf{10}, 020 (2010)
doi:10.1088/1475-7516/2010/10/020
[arXiv:1008.0260 [astro-ph.CO]].

\bibitem{Kundu:2014gxa}
N.~Kundu, A.~Shukla and S.~P.~Trivedi,
JHEP \textbf{04}, 061 (2015)
doi:10.1007/JHEP04(2015)061
[arXiv:1410.2606 [hep-th]].

\bibitem{Kundu:2015xta}
N.~Kundu, A.~Shukla and S.~P.~Trivedi,
JHEP \textbf{01}, 046 (2016)
doi:10.1007/JHEP01(2016)046
[arXiv:1507.06017 [hep-th]].

\bibitem{Bravo:2017wyw}
R.~Bravo, S.~Mooij, G.~A.~Palma and B.~Pradenas,
JCAP \textbf{05}, 024 (2018)
doi:10.1088/1475-7516/2018/05/024
[arXiv:1711.02680 [astro-ph.CO]].


\bibitem{Lifshitz:1963ps}
E.~M.~Lifshitz and I.~M.~Khalatnikov,
Adv. Phys. \textbf{12}, 185-249 (1963)
doi:10.1080/00018736300101283

\bibitem{Belinsky:1970ew}
V.~A.~Belinsky, I.~M.~Khalatnikov and E.~M.~Lifshitz,
Adv. Phys. \textbf{19}, 525-573 (1970)
doi:10.1080/00018737000101171

\bibitem{Belinskii:1972sg}
V.~A.~Belinskii, E.~M.~Lifshitz and I.~M.~Khalatnikov,
Zh. Eksp. Teor. Fiz. \textbf{62}, 1606-1613 (1972)

\bibitem{Gleyzes:2013ooa}
J.~Gleyzes, D.~Langlois, F.~Piazza and F.~Vernizzi,
JCAP \textbf{08}, 025 (2013)
doi:10.1088/1475-7516/2013/08/025
[arXiv:1304.4840 [hep-th]].

\bibitem{Fasiello:2014aqa}
M.~Fasiello and S.~Renaux-Petel,
JCAP \textbf{10}, 037 (2014)
doi:10.1088/1475-7516/2014/10/037
[arXiv:1407.7280 [astro-ph.CO]].

\bibitem{Wang:2017krj}
B.~Wang and Y.~Zhang,
Phys. Rev. D \textbf{96}, no.10, 103522 (2017)
doi:10.1103/PhysRevD.96.103522
[arXiv:1710.06641 [gr-qc]].

\bibitem{Wang:2019zhj}
B.~Wang and Y.~Zhang,
Phys. Rev. D \textbf{99}, no.12, 123008 (2019)
doi:10.1103/PhysRevD.99.123008
[arXiv:1905.03272 [gr-qc]].

\bibitem{Wang:2023nsj}
B.~Wang and Y.~Zhang,
Gen. Rel. Grav. \textbf{56}, no.2, 29 (2024)
doi:10.1007/s10714-024-03214-y
[arXiv:2307.08261 [gr-qc]].

\bibitem{Kobayashi:2011nu}
T.~Kobayashi, M.~Yamaguchi and J.~Yokoyama,
Prog. Theor. Phys. \textbf{126}, 511-529 (2011)
doi:10.1143/PTP.126.511
[arXiv:1105.5723 [hep-th]].

\bibitem{Kobayashi:2015gga}
T.~Kobayashi, M.~Yamaguchi and J.~Yokoyama,
JCAP \textbf{07}, 017 (2015)
doi:10.1088/1475-7516/2015/07/017
[arXiv:1504.05710 [hep-th]].

\bibitem{Ageeva:2021yik}
Y.~Ageeva, P.~Petrov and V.~Rubakov,
Phys. Rev. D \textbf{104}, no.6, 063530 (2021)
doi:10.1103/PhysRevD.104.063530
[arXiv:2104.13412 [hep-th]].

\bibitem{Martin:2013nzq}
J.~Martin, C.~Ringeval, R.~Trotta and V.~Vennin,
JCAP \textbf{03}, 039 (2014)
doi:10.1088/1475-7516/2014/03/039
[arXiv:1312.3529 [astro-ph.CO]].

\bibitem{Ganz:2022zgs}
A.~Ganz, P.~Martens, S.~Mukohyama and R.~Namba,
JCAP \textbf{04}, 060 (2023)
doi:10.1088/1475-7516/2023/04/060
[arXiv:2212.13561 [gr-qc]].

\bibitem{Frolovsky:2023xid}
D.~Frolovsky and S.~V.~Ketov,
Astronomy \textbf{2}, no.1, 47-57 (2023)
doi:10.3390/astronomy2010005
[arXiv:2302.06153 [astro-ph.CO]].

\bibitem{Choudhury:2024kjj}
S.~Choudhury, K.~Dey, S.~Ganguly, A.~Karde, S.~K.~Singh and P.~Tiwari,
Eur. Phys. J. C \textbf{85}, no.4, 472 (2025)
doi:10.1140/epjc/s10052-025-14176-z
[arXiv:2409.18983 [astro-ph.CO]].

\bibitem{Ganz:2024ihb}
A.~Ganz, P.~Martens, S.~Mukohyama and R.~Namba,
[arXiv:2407.02882 [gr-qc]].

\bibitem{Tristram:2021tvh}
M.~Tristram, A.~J.~Banday, K.~M.~G{\'o}rski, R.~Keskitalo, C.~R.~Lawrence, K.~J.~Andersen, R.~B.~Barreiro, J.~Borrill, L.~P.~L.~Colombo and H.~K.~Eriksen, \textit{et al.}
Phys. Rev. D \textbf{105}, no.8, 083524 (2022)
doi:10.1103/PhysRevD.105.083524
[arXiv:2112.07961 [astro-ph.CO]].

\bibitem{Armendariz-Picon:1999hyi}
C.~Armendariz-Picon, T.~Damour and V.~F.~Mukhanov,
Phys. Lett. B \textbf{458}, 209-218 (1999)
doi:10.1016/S0370-2693(99)00603-6
[arXiv:hep-th/9904075 [hep-th]].

\bibitem{Garriga:1999vw}
J.~Garriga and V.~F.~Mukhanov,
Phys. Lett. B \textbf{458}, 219-225 (1999)
doi:10.1016/S0370-2693(99)00602-4
[arXiv:hep-th/9904176 [hep-th]].

\bibitem{Peng:2016yvb}
Z.~P.~Peng, J.~N.~Yu, X.~M.~Zhang and J.~Y.~Zhu,
Phys. Rev. D \textbf{94}, no.10, 103531 (2016)
doi:10.1103/PhysRevD.94.103531
[arXiv:1611.02789 [gr-qc]].

\bibitem{Cai:2017dyi}
Y.~Cai and Y.~S.~Piao,
JHEP \textbf{09}, 027 (2017)
doi:10.1007/JHEP09(2017)027
[arXiv:1705.03401 [gr-qc]].

\bibitem{Ye:2019sth}
G.~Ye and Y.~S.~Piao,
Phys. Rev. D \textbf{99}, no.8, 084019 (2019)
doi:10.1103/PhysRevD.99.084019
[arXiv:1901.08283 [gr-qc]].

\bibitem{Mironov:2019qjt}
S.~Mironov, V.~Rubakov and V.~Volkova,
Phys. Rev. D \textbf{100}, no.8, 083521 (2019)
doi:10.1103/PhysRevD.100.083521
[arXiv:1905.06249 [hep-th]].

\bibitem{Ilyas:2020qja}
A.~Ilyas, M.~Zhu, Y.~Zheng, Y.~F.~Cai and E.~N.~Saridakis,
JCAP \textbf{09}, 002 (2020)
doi:10.1088/1475-7516/2020/09/002
[arXiv:2002.08269 [gr-qc]].

\bibitem{Zhu:2021whu}
M.~Zhu, A.~Ilyas, Y.~Zheng, Y.~F.~Cai and E.~N.~Saridakis,
JCAP \textbf{11}, no.11, 045 (2021)
doi:10.1088/1475-7516/2021/11/045
[arXiv:2108.01339 [gr-qc]].

\bibitem{Mironov:2024pjj}
S.~Mironov and V.~Volkova,
Int. J. Mod. Phys. A \textbf{39}, no.35, 2443011 (2024)
doi:10.1142/S0217751X24430115
[arXiv:2404.06297 [gr-qc]].

\bibitem{Zumalacarregui:2013pma}
M.~Zumalac{\'a}rregui and J.~Garc{\'\i}a-Bellido,
Phys. Rev. D \textbf{89}, 064046 (2014)
doi:10.1103/PhysRevD.89.064046
[arXiv:1308.4685 [gr-qc]].

\bibitem{Ageeva:2018lko}
Y.~A.~Ageeva, O.~A.~Evseev, O.~I.~Melichev and V.~A.~Rubakov,
EPJ Web Conf. \textbf{191}, 07010 (2018)
doi:10.1051/epjconf/201819107010
[arXiv:1810.00465 [hep-th]].

\bibitem{Ageeva:2020buc}
Y.~Ageeva, P.~Petrov and V.~Rubakov,
JHEP \textbf{12}, 107 (2020)
doi:10.1007/JHEP12(2020)107
[arXiv:2009.05071 [hep-th]].

\bibitem{Ageeva:2020gti}
Y.~Ageeva, O.~Evseev, O.~Melichev and V.~Rubakov,
Phys. Rev. D \textbf{102}, no.2, 023519 (2020)
doi:10.1103/PhysRevD.102.023519
[arXiv:2003.01202 [hep-th]].

\bibitem{Rubakov:2022fqk}
V.~A.~Rubakov and C.~Wetterich,
Symmetry \textbf{14}, no.12, 2557 (2022)
doi:10.3390/sym14122557
[arXiv:2210.11198 [gr-qc]].

\bibitem{Kobayashi:2010cm}
T.~Kobayashi, M.~Yamaguchi and J.~Yokoyama,
Phys. Rev. Lett. \textbf{105}, 231302 (2010)
doi:10.1103/PhysRevLett.105.231302
[arXiv:1008.0603 [hep-th]].

\bibitem{Ageeva:2022byg}
Y.~A.~Ageeva and P.~K.~Petrov,
Phys. Usp. \textbf{66}, no.11, 1134-1141 (2023)
doi:10.3367/UFNe.2022.11.039259
[arXiv:2206.03516 [hep-th]].

\bibitem{Grojean:2007zz}
C.~Grojean,
Phys. Usp. \textbf{50}, 1-35 (2007)
doi:10.1070/PU2007v050n01ABEH006157

\bibitem{Erickson:2003zm}
J.~K.~Erickson, D.~H.~Wesley, P.~J.~Steinhardt and N.~Turok,
Phys. Rev. D \textbf{69}, 063514 (2004)
doi:10.1103/PhysRevD.69.063514
[arXiv:hep-th/0312009 [hep-th]].

\bibitem{Komatsu:2001rj}
E.~Komatsu and D.~N.~Spergel,
Phys. Rev. D \textbf{63}, 063002 (2001)
doi:10.1103/PhysRevD.63.063002
[arXiv:astro-ph/0005036 [astro-ph]].

\bibitem{Bartolo:2001cw}
N.~Bartolo, S.~Matarrese and A.~Riotto,
Phys. Rev. D \textbf{65}, 103505 (2002)
doi:10.1103/PhysRevD.65.103505
[arXiv:hep-ph/0112261 [hep-ph]].

\bibitem{Creminelli:2003iq}
P.~Creminelli,
JCAP \textbf{10}, 003 (2003)
doi:10.1088/1475-7516/2003/10/003
[arXiv:astro-ph/0306122 [astro-ph]].

\bibitem{Lyth:2005du}
D.~H.~Lyth and Y.~Rodriguez,
Phys. Rev. D \textbf{71}, 123508 (2005)
doi:10.1103/PhysRevD.71.123508
[arXiv:astro-ph/0502578 [astro-ph]].

\bibitem{Lyth:2005fi}
D.~H.~Lyth and Y.~Rodriguez,
Phys. Rev. Lett. \textbf{95}, 121302 (2005)
doi:10.1103/PhysRevLett.95.121302
[arXiv:astro-ph/0504045 [astro-ph]].

\bibitem{Byrnes:2006vq}
C.~T.~Byrnes, M.~Sasaki and D.~Wands,
Phys. Rev. D \textbf{74}, 123519 (2006)
doi:10.1103/PhysRevD.74.123519
[arXiv:astro-ph/0611075 [astro-ph]].

\bibitem{Gangui:1993tt}
A.~Gangui, F.~Lucchin, S.~Matarrese and S.~Mollerach,
Astrophys. J. \textbf{430}, 447-457 (1994)
doi:10.1086/174421
[arXiv:astro-ph/9312033 [astro-ph]].

\bibitem{Verde:1999ij}
L.~Verde, L.~M.~Wang, A.~Heavens and M.~Kamionkowski,
Mon. Not. Roy. Astron. Soc. \textbf{313}, L141-L147 (2000)
doi:10.1046/j.1365-8711.2000.03191.x
[arXiv:astro-ph/9906301 [astro-ph]].

\bibitem{Creminelli:2005hu}
P.~Creminelli, A.~Nicolis, L.~Senatore, M.~Tegmark and M.~Zaldarriaga,
JCAP \textbf{05}, 004 (2006)
doi:10.1088/1475-7516/2006/05/004
[arXiv:astro-ph/0509029 [astro-ph]].

\bibitem{Akama:2019qeh}
S.~Akama, S.~Hirano and T.~Kobayashi,
Phys. Rev. D \textbf{101}, no.4, 043529 (2020)
doi:10.1103/PhysRevD.101.043529
[arXiv:1908.10663 [gr-qc]].

\bibitem{Komatsu:2000vy}
E.~Komatsu and D.~N.~Spergel,
[arXiv:astro-ph/0012197 [astro-ph]].

\bibitem{Komatsu:2001ysk}
E.~Komatsu,
[arXiv:astro-ph/0206039 [astro-ph]].

\bibitem{Gorbunov:2011zzc}
D.~S.~Gorbunov and V.~A.~Rubakov,
doi:10.1142/7873
\end{thebibliography}
\end{document}